\documentclass[twocolumn]{aastex61}
\usepackage{longtable}
\usepackage{diagbox}
\usepackage{scalerel}
\usepackage{graphicx}

\received{?}
\revised{?, 2022}

\submitjournal{AJ}

\shorttitle{HAeBe stars in LDN\,1667}
\shortauthors{Pereira et al.}

\begin{document}

\title{Herbig Ae/Be stars towards the dark cloud LDN\,1667\footnote
{Based on the observations made with the 2.2\,m telescope at the European
Southern Observatory (La Silla, Chile) under the agreement between Observat\'orio
Nacional (Brazil) and European Southern Observatory (ESO) and also under the
agreement between Observat\'orio Nacional (Brazil) and Max Planck Institut f\"ur 
Astronomie (MPG) and also based on observations at Centro Astron\'omico
Hispano Alem\'an (CAHA) at Calar Alto operated jointly by Instituto
de Astrof\'{\i}sica de Andaluc\'{\i}a (CSIC) and Max Planck Institut f\"ur 
Astronomie (MPG). Centro Astron\'omico Hispano en Andaluc\'{\i}a is
now operated by Instituto de Astrof\'{\i}sica de Andaluc\'{\i}a and Junta
de Andaluc\'{\i}a.}}

\author{C.~B. Pereira$^{1}$, L.~F. Miranda$^{2}$ \& W.~L.~F. Marcolino$^{3}$}
\affil{Observat\'orio Nacional/MCTIC, Rua Gen. Jos\'e Cristino, 77, 20921-400,
Rio de Janeiro, Brazil \\ e-mail: claudio@on.br \\$^{2}$Instituto de
Astrof\'{\i}sica de Andaluc\'{\i}a--CSIC, C/ Glorieta de la Astronom\'{\i}a s/n,
E-18008, Granada, Spain \\ lfm@iaa.es
\\ $^{3}$Universidade Federal do Rio de Janeiro, Observat\'orio do Valongo. Ladeira Pedro Ant\'onio, 43.CEP 20080-090,
Rio de Janeiro, Brazil \\ e-mail : wagner@astro.ufrj.br}

\received{xxx}
\revised{xxx}
\accepted{xxx}
\published{xxx}
\submitjournal{AJ}


\begin{abstract}
  
\par We report the discovery of a new emission-line object, named
SPH\,4$-$South = (GAIA EDR3 5616553300192230272), towards the
dark cloud LDN\,1667. This object came to our attention after
inspecting public images that show a faint diffuse nebula a few arcsec
southern from SPH\,4, an emission-line object previously classified as
a T\,Tauri star. We present high-resolution spectra and analyzed JHK
photometry of SPH\,4 and SPH\,4$-$South, and new narrow-band and
archival broad-band images of these objects. A comparison of the
spectra of SPH\,4 and SPH\,4$-$South with high-resolution ones of
DG\,Cir and R\, Mon, strongly suggests that SPH\,4 and SPH\,4$-$South
are Herbig Ae/Be stars. The classification of SPH\,4$-$South is
further supported by using a $k$-NN algorithm to its position in H$-$K
{\it versus} J$-$H color-color diagram. Both stars are detected
  in the four WISE bands and the WISE colors allow us to classify
  SPH\,4 as a Class\,I and SPH\,4$-$South as a Class\,II source.  We
also show that the faint nebula is most probably associated with
SPH\,4$-$South. Using published results on LDN\,1667 and the {\it Gaia
  Early Data Release 3}, we conclude that SPH\,4 is a member of
LDN\,1667. The case of SPH\,4$-$South is not clear because the
determination of its distance and proper motion could be affected by
the nebulosity around the star, although membership of SPH\,4$-$South
to LDN\,1667 cannot be ruled out.
\end{abstract}  


\section{Introduction}

\par SPH\,4 was first discovered as an H$\alpha$ emission-line object
by Schwartz, Persson, \& Hamann (1990) in the Canis Majoris region
and, in particular, towards the dark cloud LDN\,1667 (Lynds
1962). This and other 24 emission-line objects were studied using
low-resolution spectroscopy by Pereira et al. (2001; hereafter P2001)
with the aim of investigating their nature. As a result of that
investigation, 16 new Be stars and 7 new T\,Tauri stars were
identified while two objects failed to show the H$\alpha$ emission
line in the new spectra.

\par Among the seven stars classified as new T Tauri stars, it was
later realized that two of them, SPH\,4 and SPH\,17, could be in fact
Herbig AeBe (HAeBe) stars rather than T\,Tauri stars. The source of
this suspicion was the absence of the H component of the Ca\,{\sc ii}
line in their spectra. This K:H anomaly seems to be a common feature
among some HAeBe stars (Herbig et al. 2003) and is due to the presence
of the absorption of the P-Cygni profile of the H$\epsilon$ line at
3970.08\,{\AA} which obliterates the calcium line at
3968.49\,{\AA}. In Figure\,1 we illustrate this phenomenon by
comparing high-resolution UVES spectra of the HAeBe star Z\,CMa and of
the T\,Tauri star BP\,Tau. As can be seen in Figure\,1, the K and H
calcium lines are present in the spectrum of BP\,Tau while the H
calcium line is absent in the spectrum of Z\,CMa. In the
low-resolution spectra of both SPH\,4 and SPH\,17 the H calcium line
was absent, a fact that was not realized by P2001. Therefore, during a
mission in March 2016 at the ESO, SPH\,4 was re-observed, this time
with high-resolution spectroscopy, to obtain a more precise
classification of its nature.

\par Besides, inspecting the images of the stellar field in
Simbad/Aladin around SPH\,4 (already mentioned as Bran\,23 by Brand,
Blitz \& Wouterloot (1986)), we noticed a stellar-like object at
$\sim$14\,arcsec southern of SPH\,4, that seemed to be surrounded by a
faint diffuse nebula. Therefore, we included this object, hereafter
called SPH\,4$-$South, in the target list for our spectroscopic
observations, and also obtained narrow-band optical images of the
region around SPH\,4 to investigate the faint nebula. The coordinates
of SPH\,4, and SPH\,4$-$South are given in Table\,1 and have been
obtained from the {\it Gaia Early Data Release 3} (GEDR\,3, Gaia
Collaboration, Brown et al. 2020).

\par In this paper, we present the high-resolution spectra of SPH\,4
and SPH\,4$-$South, our narrow-band and public optical broad-band
images, and analyze JHK photometry of the two objects. We compare our
spectra with high-resolution ones of two well-known HAeBe stars,
DG\,Cir and R\,Mon, to classify our targets, and discuss the
association between SPH\,4, SPH\,4$-$South, the faint nebula, and
LDN\,1667.

\section{Observations}

\par The high-resolution spectra of SPH\,4 and SPH\,4--South were
obtained with the Feros (Fiberfed Extended Range Optical Spectrograph)
echelle spectrograph, Kaufer et al., 1999) of the 2.2\,m ESO telescope
at La Silla (Chile). SPH\,4 was observed in the night of March 23,
2016 with a exposure time of 3600\,s while SPH\,4--South was observed
in the nights of March 20 and 21, 2016, with exposure times of 3600
and 3000\,s, respectively. The FEROS spectral resolving power is
$R=48000$, corresponding to 2.2 pixels of $15\,\mu$m, and the
wavelength coverage goes from 4000\,{\AA} to 9200\,{\AA}. The spectra
were reduced with the {\sc midas} pipeline reduction package
consisting of the following standard steps: CCD bias correction,
flat-fielding, spectrum extraction, wavelength calibration, correction
of barycentric velocity and spectrum rectification. The spectrum of
DG\,Cir, used as a HAeBe comparison star, was also obtained with the
Feros spectrograph on May 2, 2010. Like the spectra of BP Tau and Z
CMa, the spectrum of R Mon used in this study was also obtained with
the UVES spectrograph.

\par Due to the large exposure times, the spectra of SPH\,4 and
SPH\,4$-$South presented many cosmic rays events and OH telluric
emission lines. We used the full width at half maximum (FWHM) to
distinguish cosmic rays from telluric emission lines, and a comparison
between spectra of the same object taken at different nights. For the
identification of OH telluric emission lines we used the
high-resolution night-sky emission atlas provided by Osterbrock et
al. (1995) and Hanuschik (2003).

\begin{table*}
\centering
\caption{Identification and coordinates of SPH\,4 and SPH\,4$-$South
from GEDR\,3.}
\begin{tabular}{lccc}
\hline
Star    &   ID    &  $\alpha$(2016.0)   &
$\delta$(2016.0) \\
         &         & ($^{\rm h}$ $^{\rm m}$ $^{\rm s}$) & ($^{\circ}$ $'$
$''$) \\
\hline

SPH\,4  & 5616553304497133568  & 7 24 12.756         & $-$25 49 57.91   
\\

SPH\,4$-$South & 5616553300192230272  &  7 24 13.167 & $-$25 50 10.33   
\\
\hline
\end{tabular}

\end{table*}

\par Narrow-band images of the field around SPH\,4 and SPH\,4$-$South
were obtained on 2017 November 1 with CAFOS at the 2.2\,m telescope at
the Calar Alto Observatory (Almer\'{\i}a, Spain).  The detector was a
SiTe CCD with 2048$\times$2048\,pixel$^2$ and a plate scale of
0.53\,arcsec\,pixel$^{-1}$.  We employed two filters covering the
H$\alpha$ ($\lambda$$_0$ = 6580\,{\AA}, FWHM = 100\,{\AA}) and the
[S\,{\sc ii}] ($\lambda$$_0$ = 6700\,{\AA}, FWHM = 180\,{\AA})
emission lines to obtain images with an exposure time of
2$\times$1800\,s in each filter. The spatial resolution, mainly
determined by the seeing, is $\sim$2.5\,arcsec.  The images were
cosmic ray cleaned, bias subtraction, and flat fielded using the
corresponding tasks in the {\sc midas} package.

\par To complete our set of data, we have downloaded the images of the
region around SPH\,4 and SPH\,4$-$South from the PanSTARRS1 archive
(Chambers et al. 2016; Flewelling et al., 2016) in the following
filters: g ($\lambda$$_0$ = 4881\,{\AA}, FWHM = 1256\,{\AA}), r
($\lambda$$_0$ = 6198\,{\AA}, FWHM = 1404\,{\AA}), and y
($\lambda$$_0$ = 9510\,{\AA}, FWHM = 628\,{\AA}). These images have a
spatial resolution of $\sim$2\,arcsec, somewhat better than the
narrow-band ones.

\begin{figure}
\begin{center}
\includegraphics[width=\columnwidth]{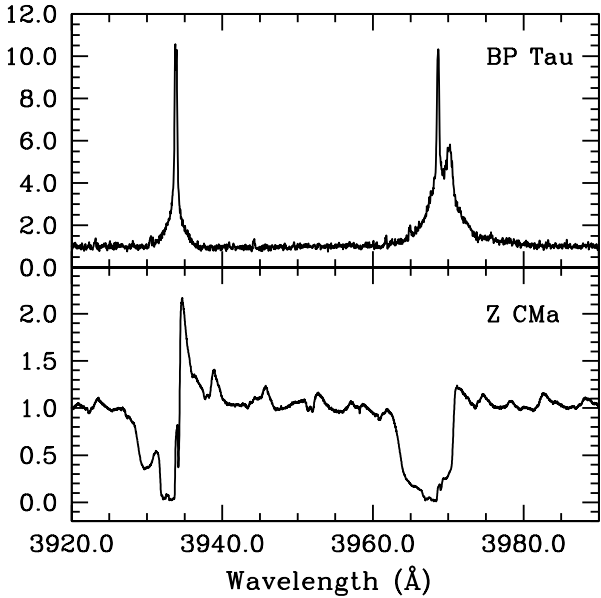}
\caption{Normalized spectra of the T Tauri star BP\,Tau and the
  HAeBe star Z\,CMa between 3920 and 3990\,{\AA}. In BP\,Tau the H calcium
  line is blended with the H$\epsilon$ 3970.08\,\AA\ line, while in Z\,CMa this line
  is not present due the P-Cyg absorption of the H$\epsilon$ line.}
\end{center}
\end{figure}

\begin{figure*}
 \begin{center} 
\includegraphics[width=170mm]{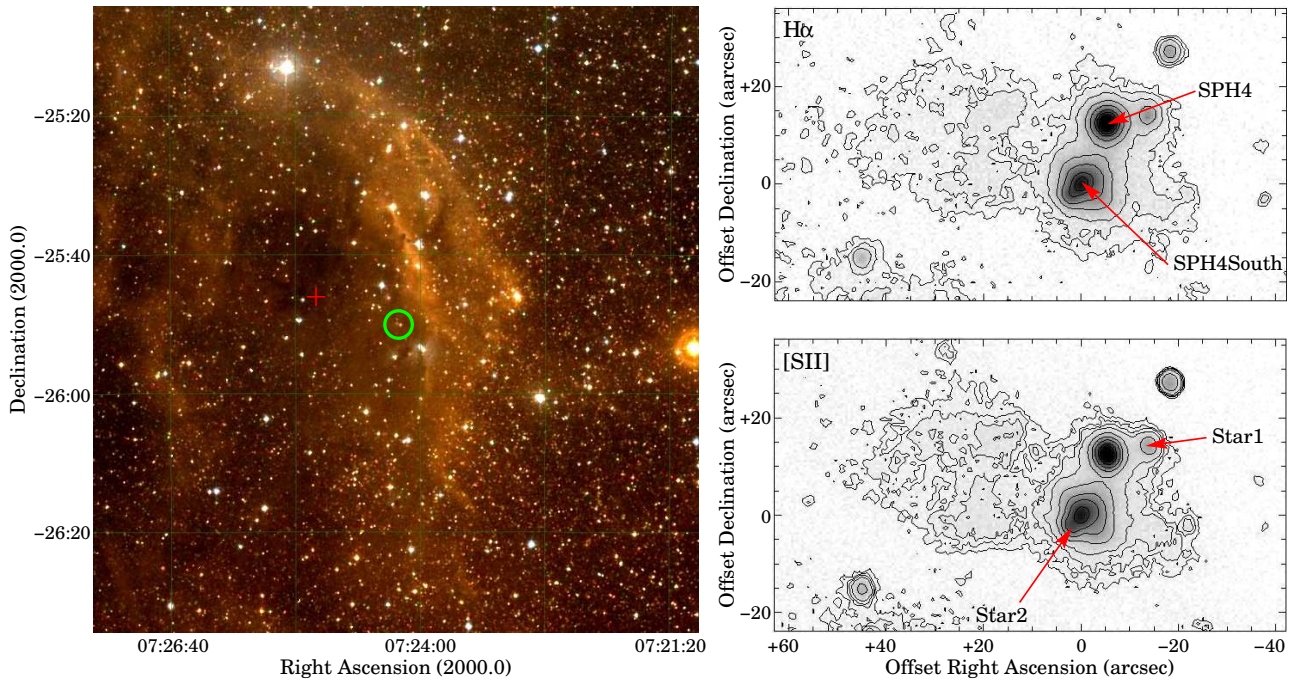}
\caption{({\it left}) Image of the LDN\,1667 dark cloud obtained from the POSS. The red cross 
  marks the center of the cloud at $\alpha$(2000) = 7$^h$ 25$^m$ 6$^s$ and $\delta$(2000) = $-$25$^{\circ}$ 46$'$ (Lynds 1962). The green
  circle marks the location of SPH\,4 and SPH\,4$-$South. ({\it right}) Grey-scale and contour
  reproductions of the H$\alpha$ and [S\,{\sc ii}] images around SPH\,4 and SPH\,4$-$South that are arrowed as well as two field stars,
  Star\,1 and Star\,2, that discussed in the text.}
\end{center}
\end{figure*}

\begin{figure}
  \begin{center}
\includegraphics[width=75mm]{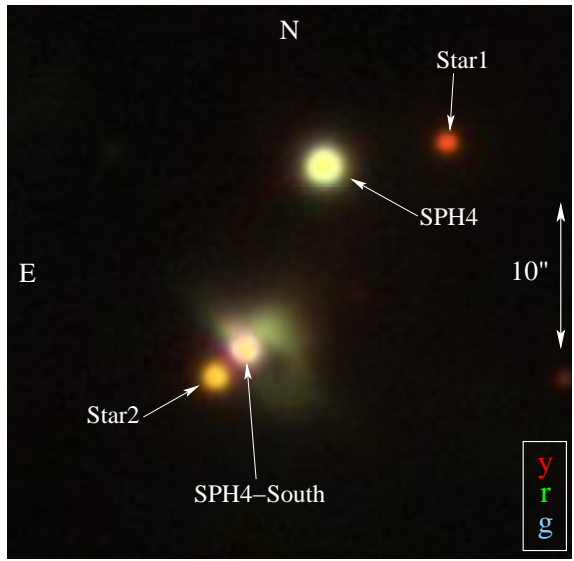}
\caption{Color composite image of SPH\,4 and SPH\,4$-$South obtained from the PanSTARRs images in the filters y, r, and g. The
  color code, scale, and orientation are indicated. SPH\,4, SPH\,4$-$South, Star\,1, and Star\,2 are arrowed (see text).}
\end{center}
\end{figure}

\section{Optical imaging}

\par Figure\,2 presents an image of LDN\,1667 obtained from the
Palomar Observatory Sky Survey (POSS) where the location of SPH\,4 and
SPH\,4$-$South is indicated, and the H$\alpha$ and [S\,{\sc ii}]
images around the two stars.  SPH\,4 and SPH\,4$-$South are located
towards the west and at $\sim$12.6\,arcmin from the center of the
cloud. They are close to bright filamentary structures. The H$\alpha$
and [S\,{\sc ii}] images show a faint bipolar nebula with the main
axis at PA$\sim$70$^{\circ}$ that crosses SPH\,4$-$South, and unequal
lobes with the eastern lobe extending up to $\sim$45\,arcsec and the
western lobe up to $\sim$23\,arcsec from the star. Some structure can
be recognized in the eastern lobe in the form of two relatively bright
regions. Figure\,3 presents a color composite image obtained from the
y, r, and g PanSTARRS1 images to show the brightest regions of the
nebula. These regions are relatively complex and show a narrow
structure towards the NE emanating from the star, bright emission
encircling a cavity towards the SW, and a bright knot towards the NW.
The images strongly suggest that the faint nebula is associated
to SPH\,4$-$South.

\par As we will see below, the spectra of SPH\,4 and SPH\,4$-$South
are mainly characterized by emission lines.  The fiber size of FEROS
is 2\,arcsec and was centered on the stars and we have not used an
off-source fiber to study the spectrum of the nebula. Therefore, it is
important to discuss whether the ``stellar'' spectra could be
contaminated by nebular emission. Emission lines due to [N\,{\sc ii}]
and, perhaps, [S\,{\sc ii}] are present in the stellar spectra and
could have a nebular origin. Balmer lines (H$\alpha$ and H$\beta$) are
strong and present a P\,Cygni profiles typical of young stars. Other
emission lines due to Fe\,{\sc ii} or Mg\,{\sc i} (see below) are
relatively strong, in many cases stronger than the [N\,{\sc ii}]
emission lines. These emission lines are very faint (if detected) in
nebulae associated to young stars and, in addition, they are not
expected from the detected nebula given its faintness. Thus, we
conclude that contamination of the stellar spectra by nebular emission
lines has not influence in the analysis and in the classification of
the stars.

\section{Line identification and the nature of SPH\,4 and SPH\,4--South}

\begin{figure}
\includegraphics[width=\columnwidth]{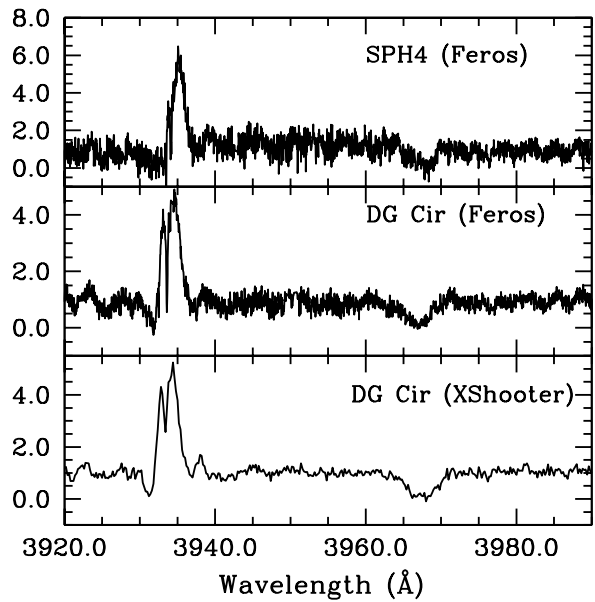}
\caption{Normalized spectra of SPH\,4 (Feros) and DG\,Cir (UVES and XShoother).
Notice the absence the H calcium emission line in both stars
and the broad absorption due the H$\epsilon$ line.}
\end{figure}

\begin{figure}
\begin{center}
  \includegraphics[width=\columnwidth]{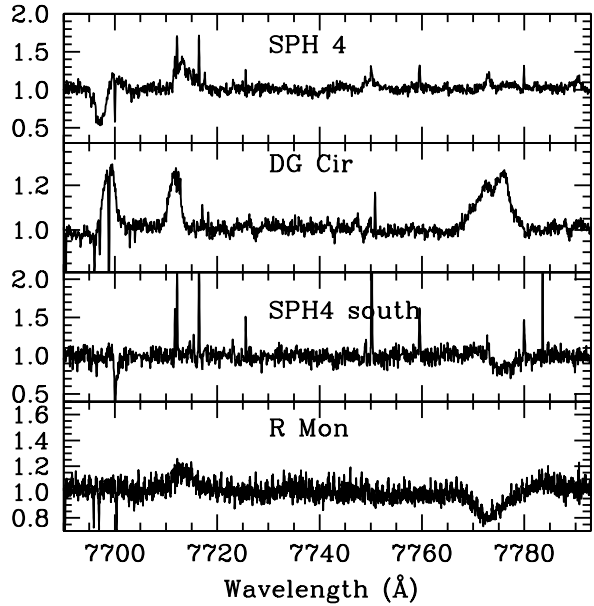}
\caption{Normalized spectra of SPH\,4, DG Cir, SPH\,4$-$South and R Mon
between 7690 and 7782\,{\AA}. Notice that the K\,{\sc i} at 7698.98\,{\AA} 
appears in absorption in SPH\,4 and SPH\,4$-$South while it appears in emission
in DG\,Cir. The Fe\,{\sc ii} at 7711.4\,{\AA} line is seen in emission
in the spectra of SPH\,4, DG\,Cir and R\,Mon. Finally the oxygen triplet
at 7771--7775\,{\AA}, in seen in absorption in SPH\,4$-$South and R\,Mon, while it appears
in emission in SPH\,4 and DG\,Cir.}
\end{center}
\end{figure}

\begin{figure}
\begin{center}  
\includegraphics[width=\columnwidth]{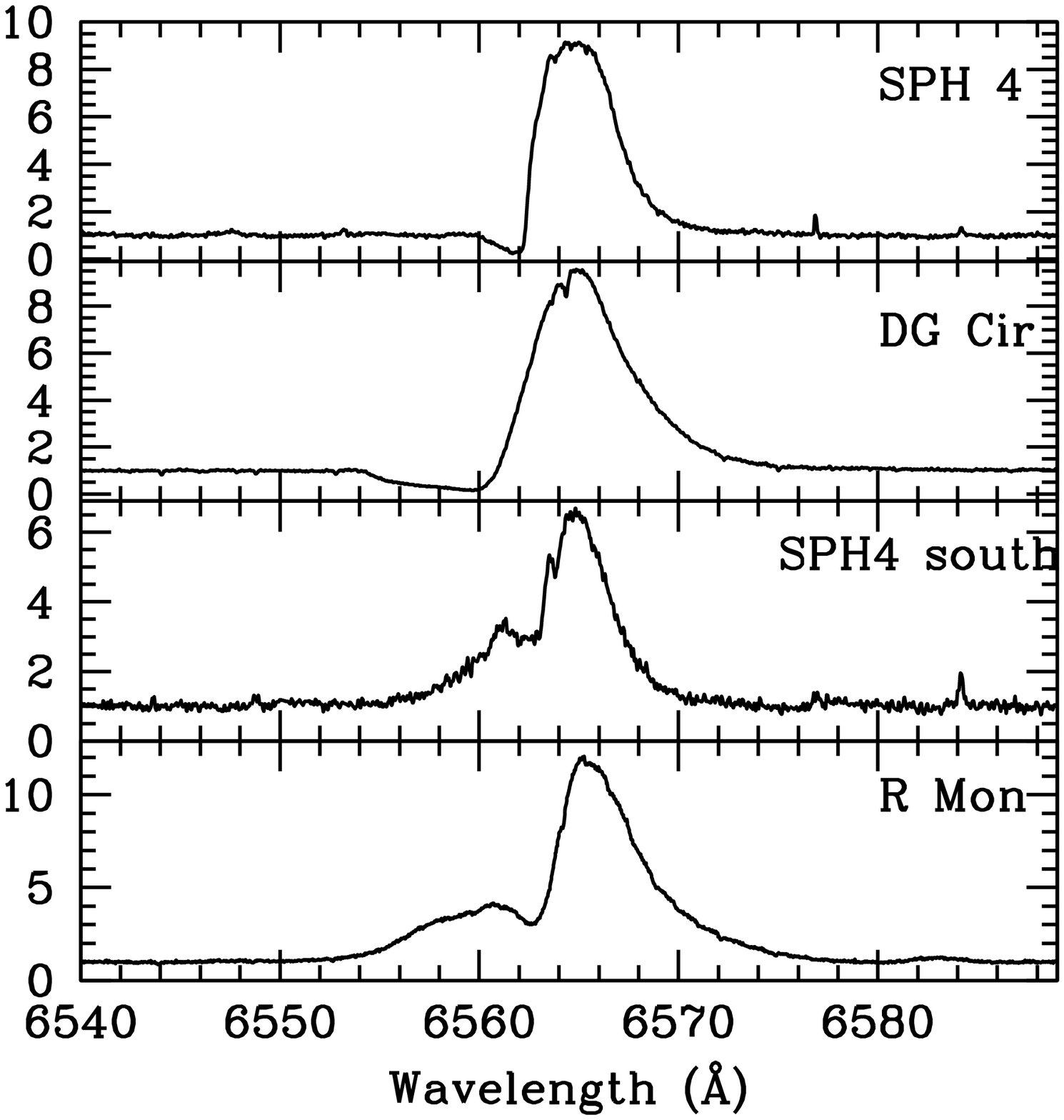}
\caption{Normalized spectra of SPH\,4, DG\,Cir, SPH\,4$-$South and R\,Mon
  around the H$\alpha$ line.
  Notice very weak [N\,{\sc ii}] emission at 6584\,{\AA} in SPH\,4 and in SPH\,4$-$South and also
  at 6548\,{\AA} in SPH\,4$-$South.}
\end{center}
\end{figure}

\par Figures\,4, 5 and 6 shows portions of the spectra of SPH\,4
and/or SPH\,4$-$South together with spectra of R\,Mon and/or DG\,Cir
that have been used as comparisons to investigate the nature of the
two emission-line stars. Figures\,10 to 22 (in the Appendix) show
additional spectral regions. 

\par Table\,5 (also in the Appendix) lists the observed wavelength,
the FWHM, the equivalent width of the identified emission lines, the
line identification, whenever possible, and the number of the Figure
showing the spectra. For some emission lines, such as H$\alpha$
(Figure 6), H$\beta$ (Figure 9), the sodium lines (Figure 16) and
[O\,{\sc i}] at 6300\,{\AA} (Figure 18), we do not provide the
equivalent width and the FWHM due to their observed complex
profiles. Besides, the emission line profiles of these these lines are
affected by circumstellar absorption and/or emission as well as by
telluric absorption and/or emission.  We note that very faint [N\,{\sc
    ii}] and, perhaps, [S\,{\sc ii}] emission lines are detected in
both stars (Figure\,6).

\par SPH\,4 presents a rich emission line spectrum, mainly due to
Fe\,{\sc ii} lines, as earlier shown in the low-resolution spectra
presented by P2001. The spectrum of this star shows remarkable
similarities with that of DG\,Cir, as can be recognized in the
figures.  Particularly noticeable is the absence of the H calcium
line, which is in agreement with that observed in DG\,Cir and other
HAeBe stars (Figures 1 and 4). The Balmer emission lines H$\beta$ and
H$\alpha$ present a P\,Cygni profile that is very similar to that
observed in DG\,Cir. From these similarities we conclude that SPH\,4
is a HAeBe star, rather than a T\,Tauri star. SPH\,4$-$South does not
exhibit a rich emission-line spectrum. In general, the spectrum of
SPH\,4$-$South is similar to that of R\,Mon, although we did not
detected any emission in the region of the H and K calcium lines. The
spectra suggest that SPH\,4$-$South may also be a HAeBe star. Alike
SPH\,4 and DG\,Cir, R\,Mon is considered as a ``continuum star'' by
Hern\'andez et al. (2004) due to a high nonphotospheric continuum
veiling and, hence, the absence of some photospheric absorption
lines. A similar classification may hold from SPH\,4$-$South. Yet, a
few absorption lines could be identified such as Mg\,{\sc i} at
5183.62\,{\AA} in the spectrum of SPH\,4, (Figure\,12), K\,{\sc i} at
7698.98\,{\AA} in the spectra of SPH\,4 and SPH\,4$-$South and the
oxygen triplet at 7771$-$7775\,{\AA} in the spectra of SPH\,4$-$South
and R\,Mon (Figure\,5).

\par From well defined emission lines, we obtain (LSR) radial
velocities of +15.2$\pm$1.3 and $-$32.3$\pm$0.9\,km\,s$^{-1}$ for
SPH\,4 and SPH\,4$-$South, respectively. We have tried to measure the
stellar velocity from the few observed absorption lines but the
faintness or broadening of these lines does not provide reliable
results.

\section{SPH\,4 and SPH\,4$-$South in the H$-$K {\it versus} J$-$H diagram and
in the WISE bands}

\par In Figure\,7 we examine the positions of SPH\,4 and
SPH\,4$-$South in the H$-$K {\it versus} J$-$H diagram based on
  data given in Table 2 and compare them with a sample of T\,Tauri
and HAeBe stars. In addition, we also insert in the diagram the
objects classified as T\,Tauri stars by P2001. Based in the infrared
diagram, it is difficult to classify SPH\,4 and SPH\,4$-$South either
as HAeBe or T\,Tauri star since they both lie in between the region
occupied by these two groups of stars. SPH\,17 previously classified
as T\,Tauri star by P2001, occupies the region of HAeBe objects. In
fact, inspecting the low-resolution spectrum of SPH\,17 (see P2001),
we see that it is very similar to the low-resolution spectrum of
SPH\,4 and, in addition, the H calcium line is absent (see
above). Therefore, we re-classify SPH\,17 as HAeBe star. The other
pre-main sequence stars SPH\,1, SPH\,6, SPH\,21, SPH\,22, and SPH\,24,
only analyzed with low-resolution spectroscopy, may still be
considered candidate T\,Tauri stars although some of them lie in
between the region occupied by the HAeBe and T\,Tauri stars.

\par Figure\,8 presents another version of the H$-$K {\it versus}
J$-$H color-color diagram. To gain insight into the classification
of SPH4 and SPH4$-$South, we computed a decision surface around their
locations using a $k$-NN algorithm. With this method, each point of
the (H$-$K, J$-$H) plane was classified as T\,Tauri or HAeBe star,
based on the most common classification of its $k$-nearest
neighbors. Using cross-validation in our data set, we found that $k=25$
was the optimal value.

\begin{figure}
\begin{center}  
\includegraphics[width=\columnwidth]{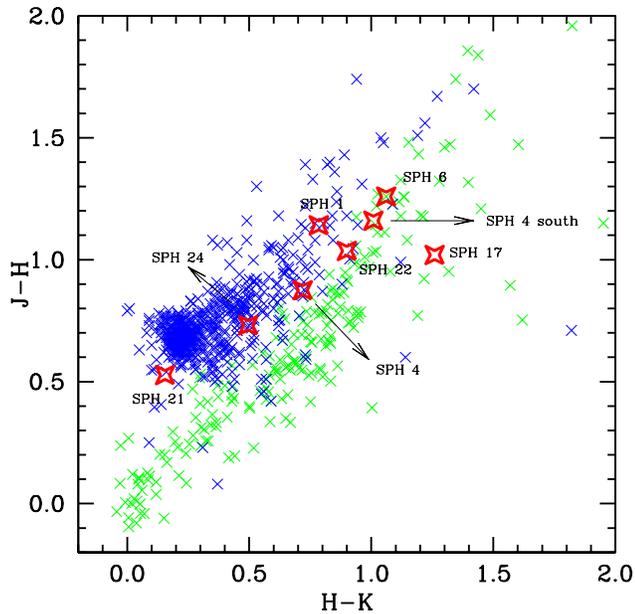}
\caption{SPH\,4, SPH\,4$-$South, and stars classified as T Tauri stars
by Pereira et al. (2001) (red stars) in the H$-$K {\it versus} J$-$H diagram.
Green crosses represent HAeBe stars with data taken from Vioque et al. (2018)
and blue crosses represent T Tauri stars with data taken from
Dahm \& Simon (2005)
and Percy et al. (2010).}
\end{center}
\end{figure}

\begin{figure}
\begin{center}  
\includegraphics[width=\columnwidth]{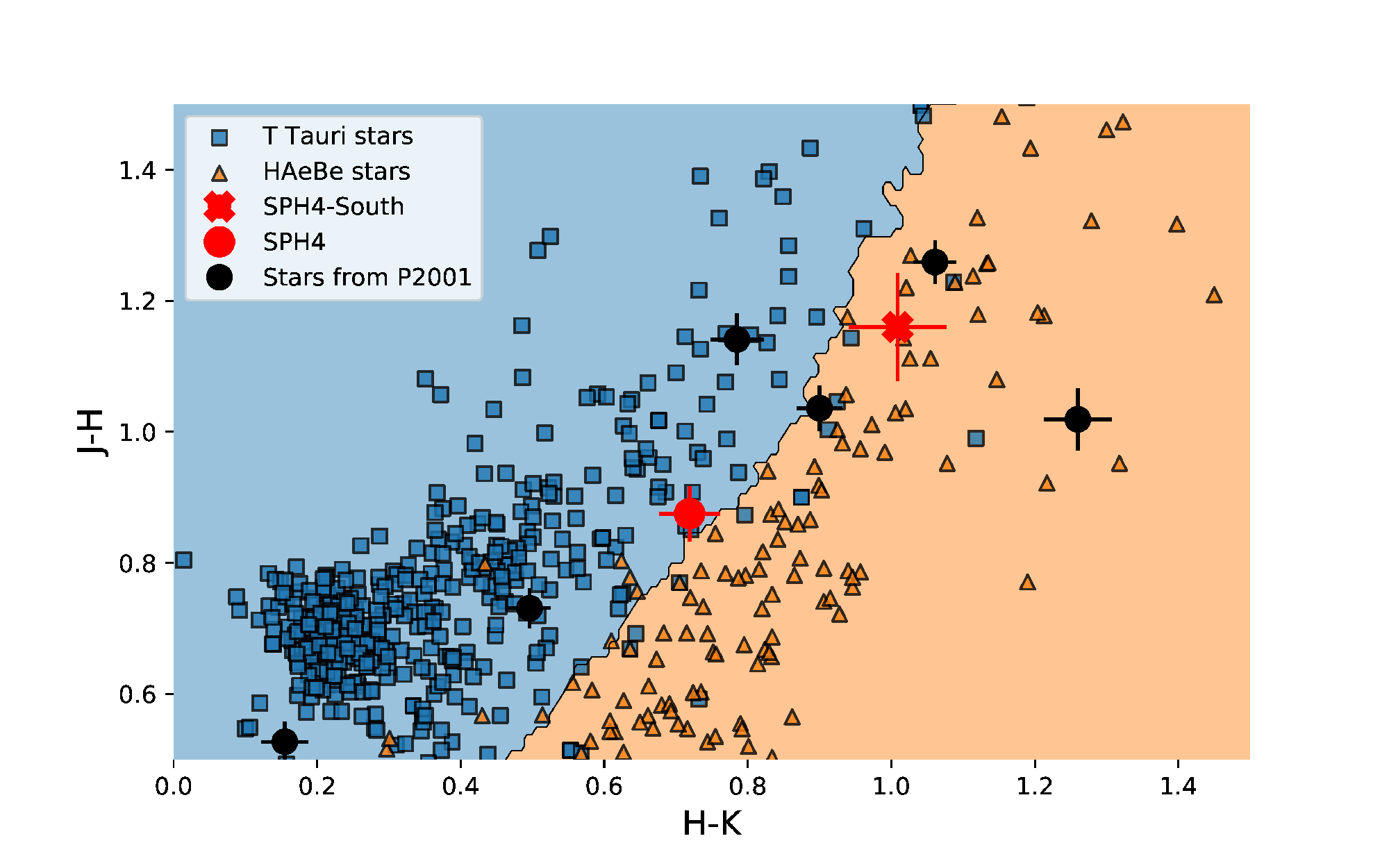}
\caption{Same as Figure\,7, but with a decision surface added. The surface
(i.e., colored regions) was computed with the $k$-NN method (see text).
Note that SPH\,4 falls close to the frontier between T\,Tauri and HAeBe stars
and SPH4$-$South is on the HAeBe side. SPH\,17 is the object with the
highest H-K value among the sample of P2001.}
\end{center}
\end{figure}

\par Overall, the T\,Tauri and HAeBe stars were well
separated. SPH4-south is on the HAeBe side and SPH4 falls very close
to the frontier between T\,Tauri and HAeBe stars. A definitive
classification for the objects close to the frontier is hard to
establish because there is some overlap in the observed
colors. Without the additional information provided by their spectra,
their classification is risky. Nevertheless, Figure 8 at least
indicates that the classification of SPH4$-$South as an HAeBe star is
indeed statistically possible.

\par The accuracy achieved with the kNN method was high. The number of
misclassified T-Tauri stars is 26, which is only about 6\% of the
total T-Tauri sample. For the HAeBe stars, there were 14
misclassifications, which is about 7\% of the HAeBe sample. In
Figure\,8, each of these objects are in the wrong side of the computed
frontier.  We also display in Figure\,8 the error bars for the H$-$K
and J$-$H colors of SPH\,4 and SPH\,4\- South, computed from
individual magnitude errors. The error bars for SPH\,4 are relatively
small and barely cross the frontier. Analogously, SPH\,4-South remains
on the HAeBe side, despite the larger color uncertainties.

\begin{table*}
\centering
\caption{JHK photometry for the stars studied in this work with data taken
from Cutri et al. (2003).}
\begin{tabular}{lcccccc} 
\hline
Star        &    J    &    e   &    H    &    e   &    K    &    e    \\\hline
SPH\,1      &  13.102 &  0.027 &  11.961 &  0.029 &  11.176 &  0.023\\
SPH\,4      &  12.392 &  0.028 &  11.517 &  0.031 &  10.798 &  0.029\\
SPH\,6      &  11.610 &  0.024 &  10.351 &  0.023 &   9.290 &  0.019\\
SPH\,17     &  12.802 &   ---  &  11.783 &  0.037 &  10.523 &  0.029\\
SPH\,21     &  12.884 &  0.023 &  12.357 &  0.021 &  12.202 &  0.025\\
SPH\,22     &  13.181 &  0.027 &  12.145 &  0.022 &  11.245 &  0.023\\
SPH\,24     &  12.608 &  0.022 &  11.877 &  0.021 &  11.381 &  0.021\\
SPH\,4$-$South &  12.254 &  0.056 &  11.094 &  0.061 &  10.085 &  0.030\\\hline
\end{tabular}
\end{table*}

\par To further investigate the nature of the two stars, we have
followed the method by Fischer et al. (2016) who used the {\it
  Wide-field Infrared Survey Explorer (WISE)} and the W1$-$W2 and
W2$-$W3 colors to classify Class\,I and Class\,II sources in the Canis
Majoris region (see also Lada 1987). Table\,3 lists the $WISE$
magnitudes of the two stars that are detected in the four bands, being
SPH\,4 particularly bright in W4. According Figure\,2 in Fischer et
al. (2016), SPH\,4 may be classified as a Class\,I source and
SPH\,4$-$South as a Class\,II one. The infrared emission of Class\,I
sources is dominated by a dusty circumstellar envelope while a dusty
disk dominates the infrared emission of Class\,II sources that are
expected to be more evolved than Class\,I ones.

\begin{table*}
\centering
\caption{WISE magnitudes of SPH\,4 and SPH\,4$-$South.}
\begin{tabular}{lccccccccc} 
\hline
Star        &    W1    &    e   &   W2    &    e   &    W3    &    e  &  W4     &    e \\
\hline
SPH\,4      & 9.085    & 0.023  & 7.976    &  0.019  &  5.144  & 0.014  & 2.710  &  0.018 \\  
SPH\,4$-$South & 8.746 & 0.025  & 7.845    &  0.021  & 5.827   & 0.019  & 4.033  & 0.039  \\
\hline
\end{tabular}
\end{table*}

\section{The relationship among SPH\,4, SPH\,4$-$South and LDN\,1667}
 
As already mentioned, SPH\,4 and SPH\,4$-$South are observed towards the
dark cloud LDN\,1667 and it is interesting to investigate their possible
membership to this cloud.

LDN\,1667 has not been studied in detail. May, Gyulbudaghian
\& Alvarez (2005, hereafter MGA05) mapped several molecular clouds
towards the Pupis--Canis Majoris region using the $^{13}$CO\,(J=1-0)
line. SPH\,4 and SPH\,4$-$South are observed towards the B molecular cloud
identified by MGA05 at $l$$\sim$239.63 and $b$$\sim$$-$4.63 that coincide with
those of LDN\,1667 at $l$$\sim$239.57 and $b$$\sim$$-$4.64. These authors obtained an
LSR radial velocity between $\sim$18.42 and $\sim$23.62\,km\,s$^{-1}$ for the B cloud, and
a kinematical distance between 1.7 and 2.1\,kpc with errors of $\pm$0.5\,kpc from their
CO observations. The kinematical distance largely differs from that determined using stars
presumably being members of the region, that results to be between $\sim$300 and $\sim$650\,pc
(see MGA05).

From the data in GEDR3, we have obtained the distances, proper motions, and
magnitudes of SPH\,4 and SPH\,4$-$South that are listed in Table\,4. In addition,
for comparison purposes, the same information has been obtained for Star\,1 and
Star\,2 that are marked in Figures\,2 and 3, and is included in Table\,4.

The distance of SPH\,4 is in agreement with the kinematical distance to the
cloud B, strongly suggesting that it is a member of LDN\,1667. Star\,1 and
Star\,2 present a similar distance and they should also be members of the cloud.
Moreover, the proper motions of SPH\,4, Star\,1, and Star\,2 present similar magnitude
and direction. We also note that the (LSR) radial velocity of SPH\,4 ($\sim$+15\,km\,s$^{-1}$) 
is similar to that of LDN\,1667, although this could be a simple chance and
the stellar velocity of SPH\,4 should be used to make a better comparison.

\par The case of SPH\,4$-$South is different. Its distance is clearly
smaller than those of the other three stars and of LDN\,1667. The
magnitude of its proper motion is very different from those of the
other three stars, although the direction coincides within the errors
(Table\,4). The radial velocity of SPH\,4$-$South
($\sim$$-$32\,km\,s$^{-1}$) is very different from that of LDN\,1667,
although this value has been derived from the emission lines and
should not necessarily represent that of the star. In general, the
data seem to suggest that SPH\,4$-$South is not related to SPH\,4 and
LDN\,1667. However, we note that the errors in the distance and proper
motion of SPH\,4$-$South are considerably larger than those associated
to the other three stars. In principle, the errors could be related to
the brightness of the star, as it can be inferred from Table\,2 for
Star\,1, Star\,2, and SPH\,4. However, Star\,1 and Star\,2 are clearly
fainter than SPH\,4$-$South and show smaller errors in distance and
proper motion than SPH\,4$-$South.  We suspect that the measurements
of SPH\,4$-$South might be affected by the presence of the nebula,
that could introduce uncertainties in the measurements of its
photocenter. Therefore, membership of SPH\,4$-$South to LDN\,1667
cannot not be ruled out from the current data and obtaining its
stellar velocity is mandatory to confirm or reject such a membership.

\begin{table*}
\centering
\caption{Distance, proper motion, and magnitudes of SPH\,4, SPH\,4$-$South, Star\,1, and Star\,2 from GEDR\,3.}
\begin{tabular}{lcccccc} 
\hline
Star    &  Distance                 & Proper motion magnitude    & Proper motion PA           & G$_{Bp}$    &  G$_{G}$ & G$_{Rp}$ \\
        &  (pc)                     & (mas)            & ($^{\circ}$)   & (mag)      & (mag)    & (mag)   \\
\hline

SPH\,4  & 1387$^{+38}$\llap{$_{-34}$}  & 3.33$\pm$0.02   & 321.5$\pm$0.4       & 15.613$\pm$0.010  & 14.675$\pm$0.004 & 13.669$\pm$0.007 \\

SPH\,4$-$South & 896$^{+205}$\llap{$_{-141}$}  & 0.95$\pm$0.27   & 309$\pm$16  & 16.521$\pm$0.029   & 15.740$\pm$0.011  & 14.473$\pm$0.025 \\

Star\,1 & 1324$^{+187}$\llap{$_{-145}$}  & 3.43$\pm$0.12   & 317.3$\pm$2.2     & 19.473$\pm$0.030   & 17.947$\pm$0.003  & 16.556$\pm$0.007 \\

Star\,2 & 1381$^{+94}$\llap{$_{-83}$}   & 3.42$\pm$0.06   & 321$\pm$1          & 17.959$\pm$0.036   & 16.608$\pm$0.008  & 15.414$\pm$0.026 \\

\hline
\end{tabular}
\end{table*}

\section{Conclusions}

\par We have presented high-resolution spectra of the emission-line
star SPH\,4 and of a new emission-line object, named SPH\,4$-$South,
that is presumably associated to a faint diffuse nebula detected in
public images. We also analyzed narrow- and broad-band images to
investigate the faint nebula. By comparing the spectra of SPH\,4 and
SPH\,4$-$South with similar ones of DG\,Cir and R\,Mon, respectively,
we re-classify SPH\,4 as a HAeBe star, correcting the previous
T\,Tauri classification, and also suggest an HAeBe classification for
SPH\,4$-$South. An analysis of the H$-$K {\it versus} J$-$H
color-color diagram by means of a $k$-NN algorithm provides support
for the HAeBe classification of SPH\,4$-$South. he two stars are
detected in the four WISE bands and their W1$-$W2 and W2$-$W3 colors
suggest a Class\,I and Class\,II classification for SPH\,4 and
SPH\,4$-$South, respectively. The faint nebula appears associated to
SPH\,4$-$South.  Both stars are observed towards the dark cloud
LDN\,1667. Using published data on LDN\,1667 and GEDR\,3 we conclude
that SPH\,4 is a member of the cloud.  The GEDR\,3 data of
SPH\,4$-$South seems to exclude its association with
LDN\,1667. Nevertheless, these data present relatively large errors
that might be caused by the presence of the nebula, and membership of
SPH\,4$-$South to LDN\,1667 cannot be still ruled out.

\section*{Acknowledgements}

\par We thank Calar Alto Observatory for allocation of director's
discretionary time to this program. LFM acknowledges partial support
by MCIU grant AYA2017-84390-C2-1-R, co-funded with FEDER funds, and
financial support from the State Agency for Research of the Spanish
MCIU through the ``Center of Excellence Severo Ochoa'' award for the
Instituto de Astrof\'{\i}sica de Andaluc\'{\i}a (SEV-2017-0709). This
research has made use of the SIMBAD database, operated at CDS,
Strasbourg, France, and of the VizieR catalogue access tool, CDS,
Strasbourg, France (DOI: 10.26093/cds/vizier). The original
description of the VizieR service was published in Ochsenbein, Bauer
\& Marcout (2000). The Pan-STARRS1 Surveys (PS1) and the PS1 public
science archive have been made possible through contributions by the
Institute for Astronomy, the University of Hawaii, the Pan-STARRS
Project Office, the Max-Planck Society and its participating
institutes, the Max Planck Institute for Astronomy, Heidelberg and the
Max Planck Institute for Extraterrestrial Physics, Garching, The Johns
Hopkins University, Durham University, the University of Edinburgh,
the Queen's University Belfast, the Harvard-Smithsonian Center for
Astrophysics, the Las Cumbres Observatory Global Telescope Network
Incorporated, the National Central University of Taiwan, the Space
Telescope Science Institute, the National Aeronautics and Space
Administration under Grant No. NNX08AR22G issued through the Planetary
Science Division of the NASA Science Mission Directorate, the National
Science Foundation Grant No. AST-1238877, the University of Maryland,
Eotvos Lorand University (ELTE), the Los Alamos National Laboratory,
and the Gordon and Betty Moore Foundation. The Digitized Sky Surveys
were produced at the Space Telescope Science Institute (STScI) under
US Government grant NAG W-2166. The images of these surveys are based
on photographic data obtained using the Oschin Schmidt Telescope on
Palomar Mountain and the UK Schmidt Telescope. The plates were
processed into the present compressed digital form with the permission
of these institutions. The National Geographic Society – Palomar
Observatory Sky Atlas (POSS-I) was made by the California Institute of
Technology with grants from the National Geographic Society. The
second Palomar Observatory Sky Atlas (POSS-II) was made by the
California Institute of Technology with funds from the National
Science Foundation, the National Geographic Society, the Sloan
Foundation, the Samuel Oschin Foundation and the Eastman Kodak
Corporation. The Oschin Schmidt Telescope is operated by the
California Institute of Technology and Palomar Observatory.  The UK
Schmidt Telescope was operated by the Royal Observatory, Edinburgh,
with funding from the UK Science and Engineering Research Council
(later the UK Particle Physics and Astronomy Research Council), until
1988 June, and thereafter by the Anglo-Australian Observatory.
Supplemental funding for skysurvey work at the STScI is provided by
the European Southern Observatory. This publication makes use of data
products from the Wide-field Infrared Survey Explorer, which is a joint
project of the University of California, Los Angeles, and the Jet Propulsion
Laboratory/California Institute of Technology, funded by the National Aeronautics
and Space Administration.





\clearpage

\appendix


\section{Additional figures}

In this section we present additional figures that compare the spectra of SPH\,4 and SPH\,4$-$South
with those of DG\,Cir and R\,Mon, respectively.

\begin{figure}
\includegraphics[width=\columnwidth]{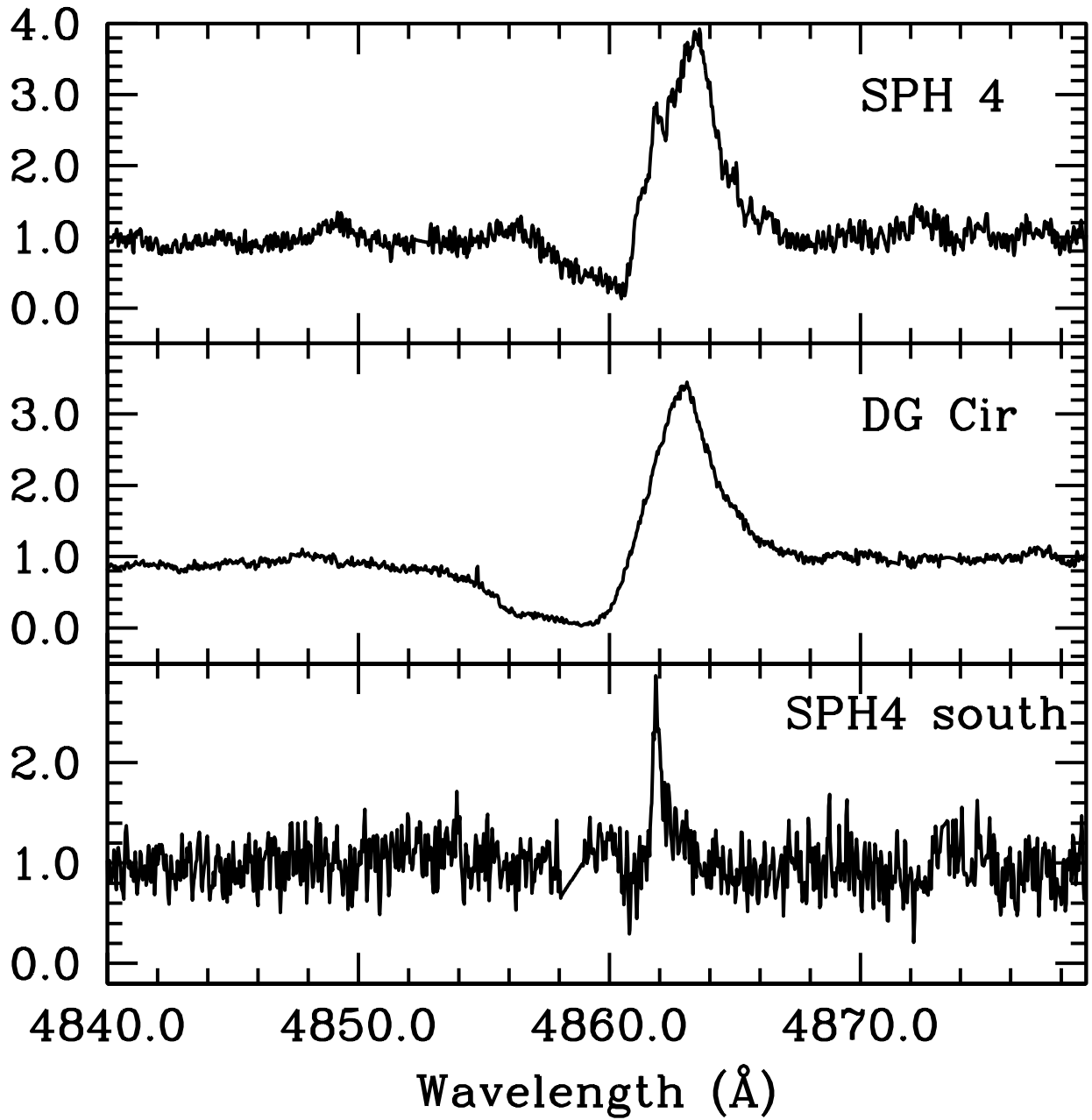}
\caption{Normalized spectra of SPH\,4, DG\,Cir, and SPH\,4$-$South
around the H$\beta$ line.}
\end{figure}

\begin{figure}
\includegraphics[width=\columnwidth]{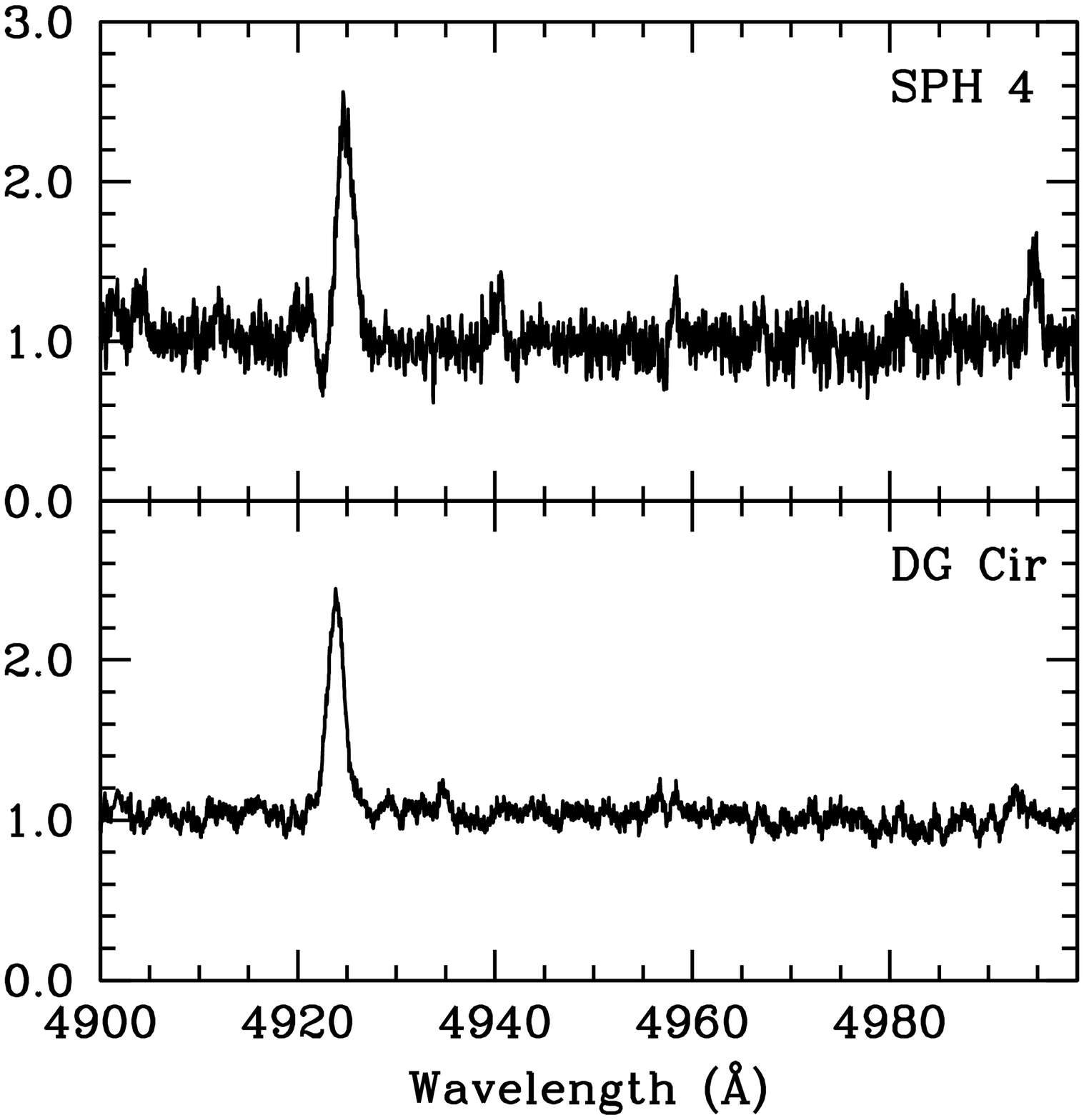}
\caption{Normalized spectra of SPH\,4 and DG\,Cir between 4900 and 5000\,{\AA}.}
\end{figure}

\begin{figure}
\includegraphics[width=\columnwidth]{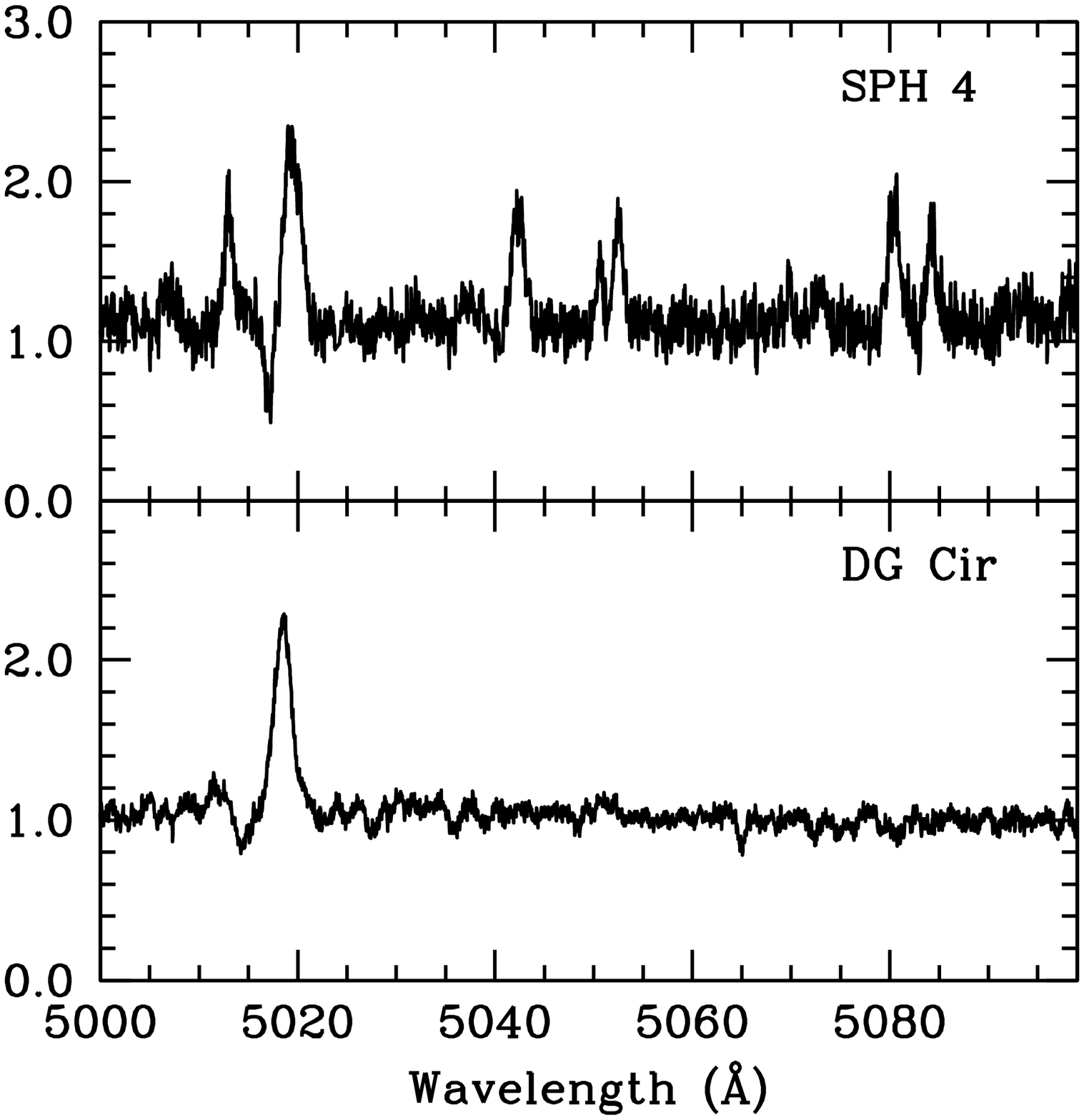}
\caption{Normalized spectra of SPH\,4 and DG\,Cir between 5000 and 5100\,{\AA}.}
\end{figure}

\begin{figure}
  \includegraphics[width=\columnwidth]{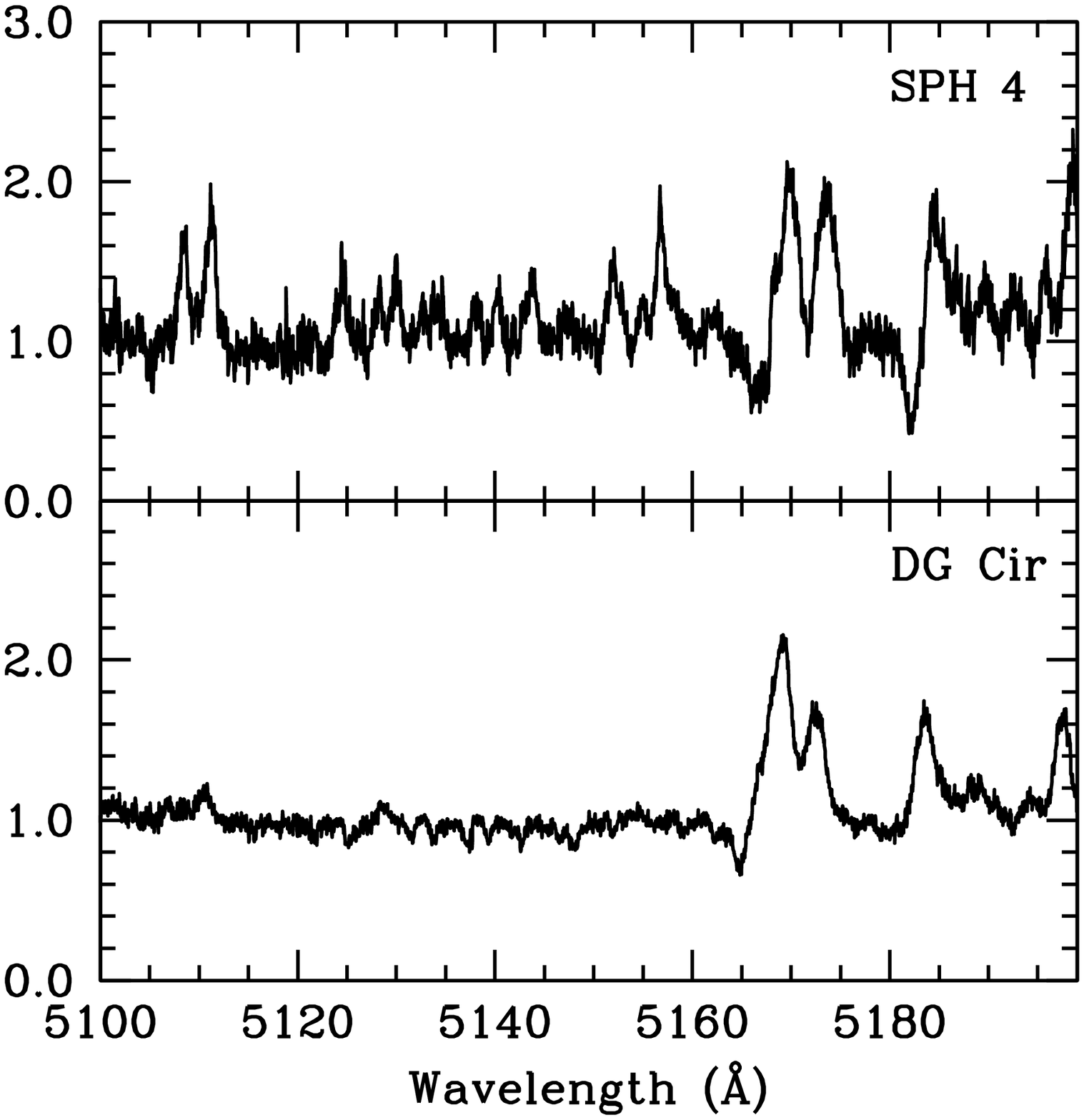}
\caption{Normalized spectra of SPH\,4 and DG\,Cir between 5100 and 5200\,{\AA}.}
\end{figure}

\begin{figure}
\includegraphics[width=\columnwidth]{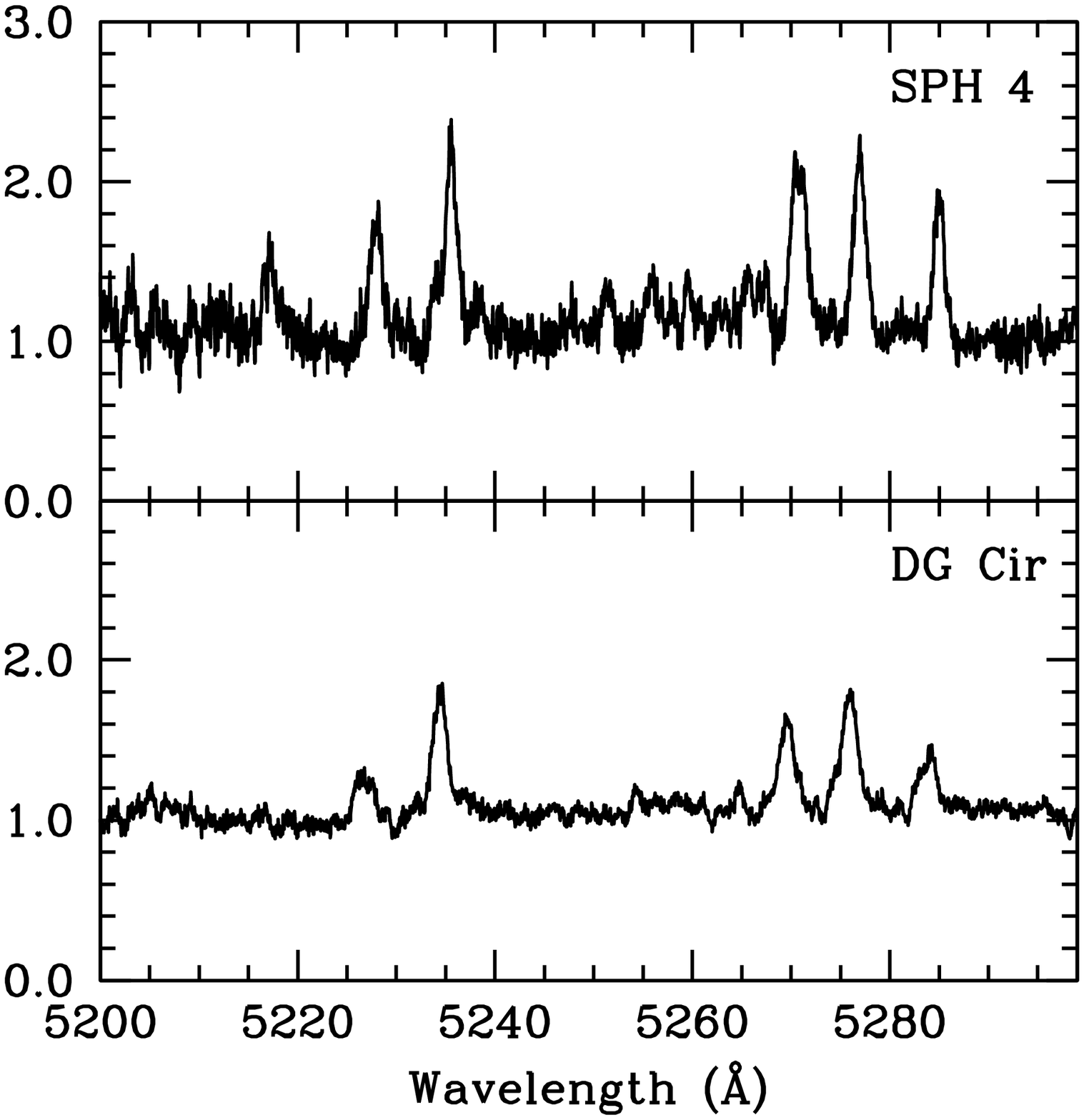}
\caption{Normalized spectra of SPH\,4 and DG\,Cir between 5200 and 5300\,{\AA}.}
\end{figure}

\begin{figure}
\includegraphics[width=\columnwidth]{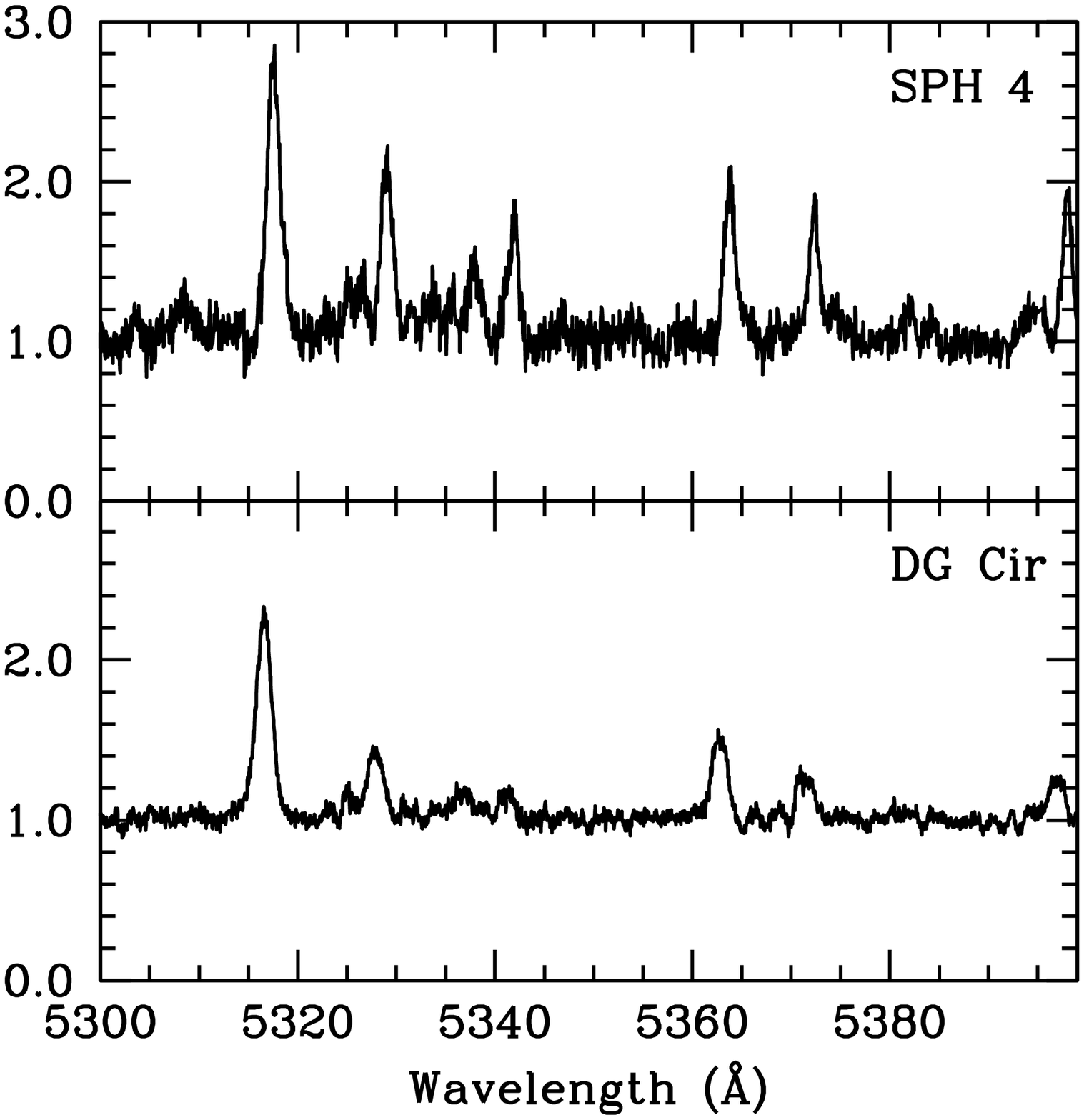}
\caption{Normalized spectra of SPH\,4 and DG\,Cir between 5300 and 5400\,{\AA}.}
\end{figure}

\begin{figure}
\includegraphics[width=\columnwidth]{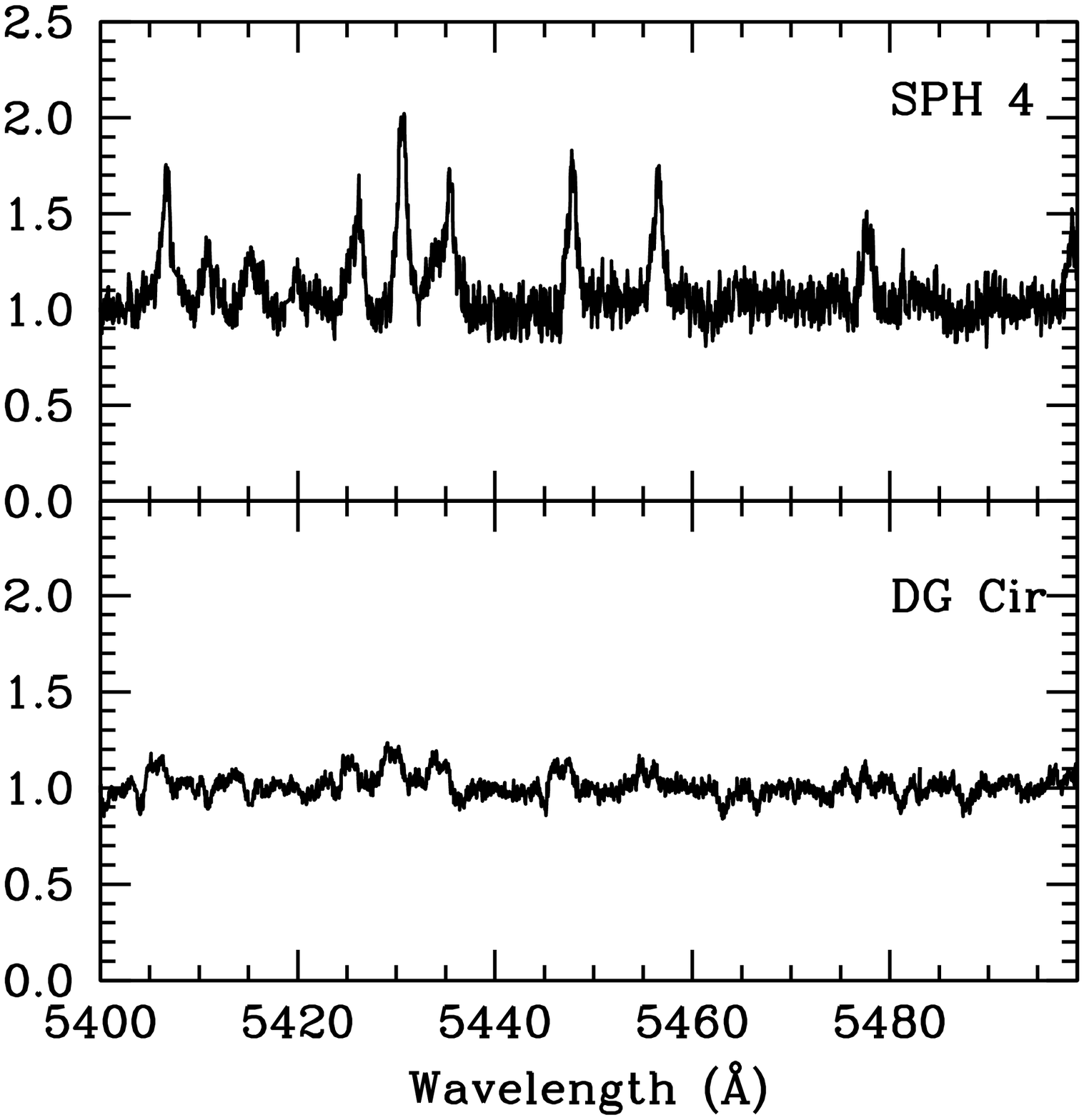}
\caption{Normalized spectra of SPH\,4 and DG\,Cir between 5400 and 5500\,{\AA}.}
\end{figure}

\begin{figure}
\includegraphics[width=\columnwidth]{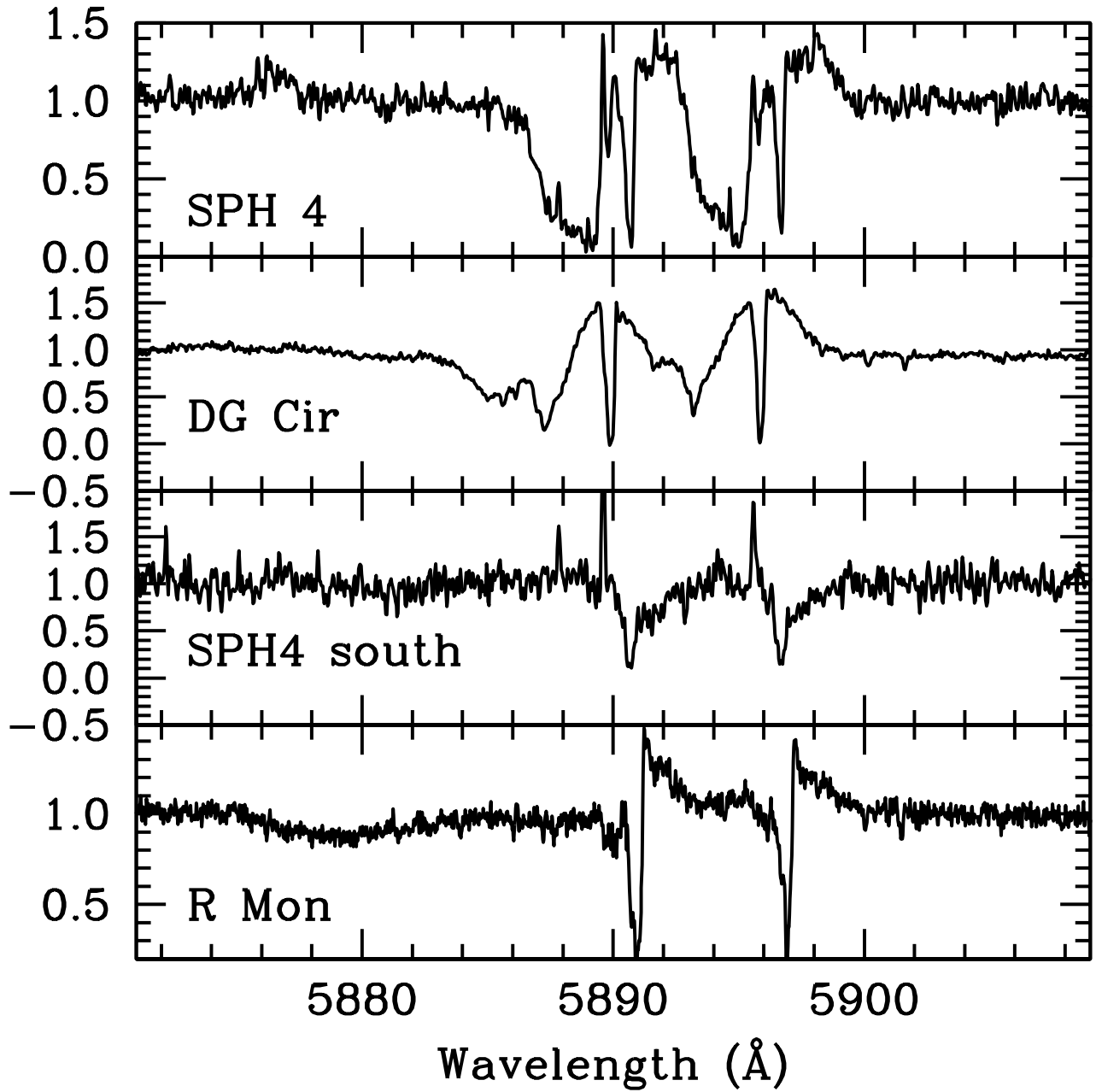}
\caption{Normalized spectra of SPH\,4, DG\,Cir SPH\,4$-$South, and R\,Mon
  around the region of Na\,{\sc i} lines. Notice the weak He\,{\sc i} emission at 5876\,{\AA}, the photospheric broad
  Na\,{\sc i} absorptions and the circunstellar narrow Na\,{\sc i} absorptions, as well as weak telluric emission Na\,{\sc i}
  lines (including weak OH emission at $\sim$5888\,{\AA}) in the spectrum of SPH\,4. In SPH\,4$-$South we notice the
  circunstellar narrow Na\,{\sc i} absorptions in the broad absorption profiles.}
\end{figure}

\begin{figure}
\includegraphics[width=\columnwidth]{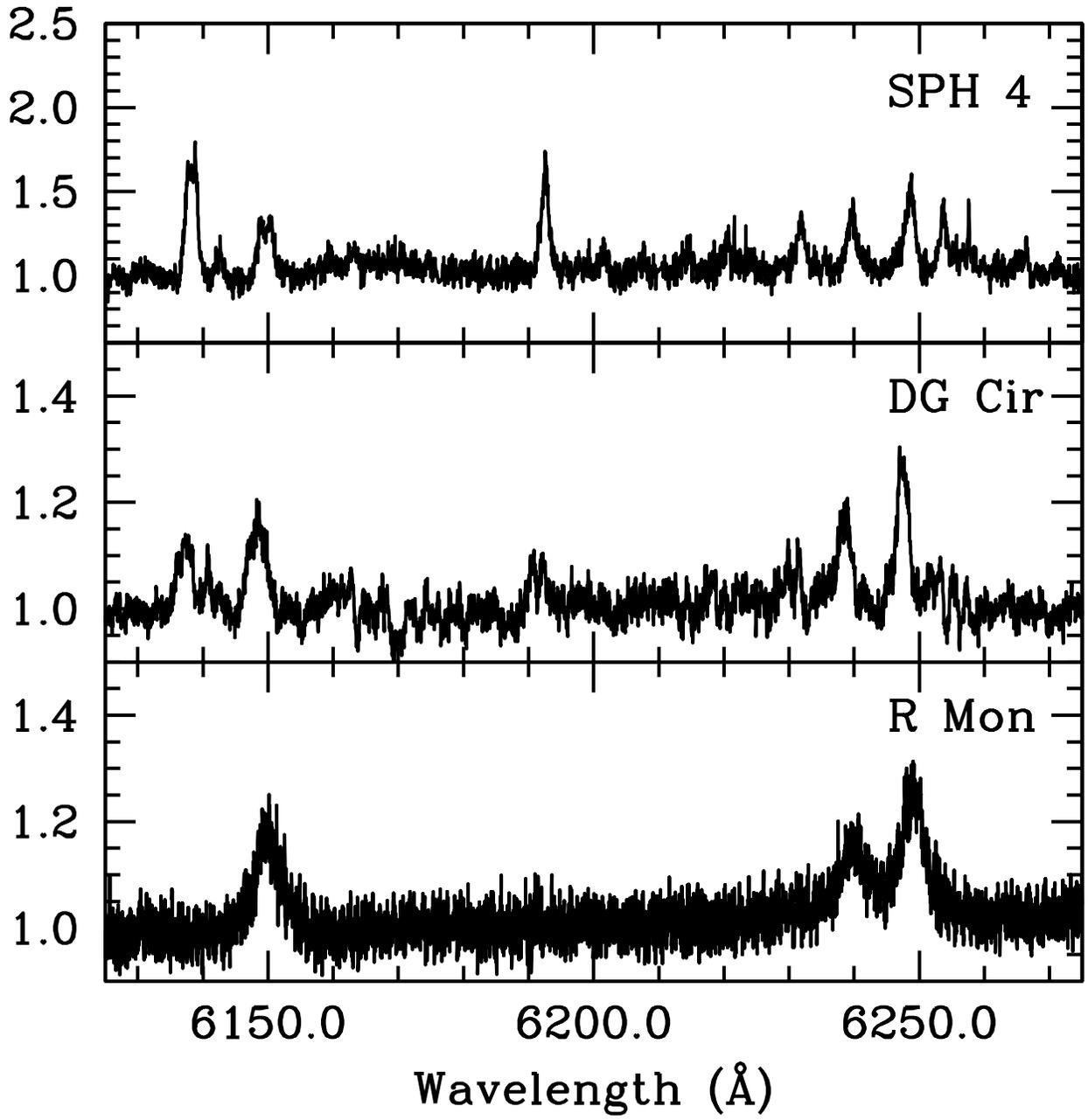}
\caption{Normalized spectra of SPH\,4, DG\,Cir, and R\,Mon between 6125 and 6275\,{\AA}.}
\end{figure}

\begin{figure}
\includegraphics[width=\columnwidth]{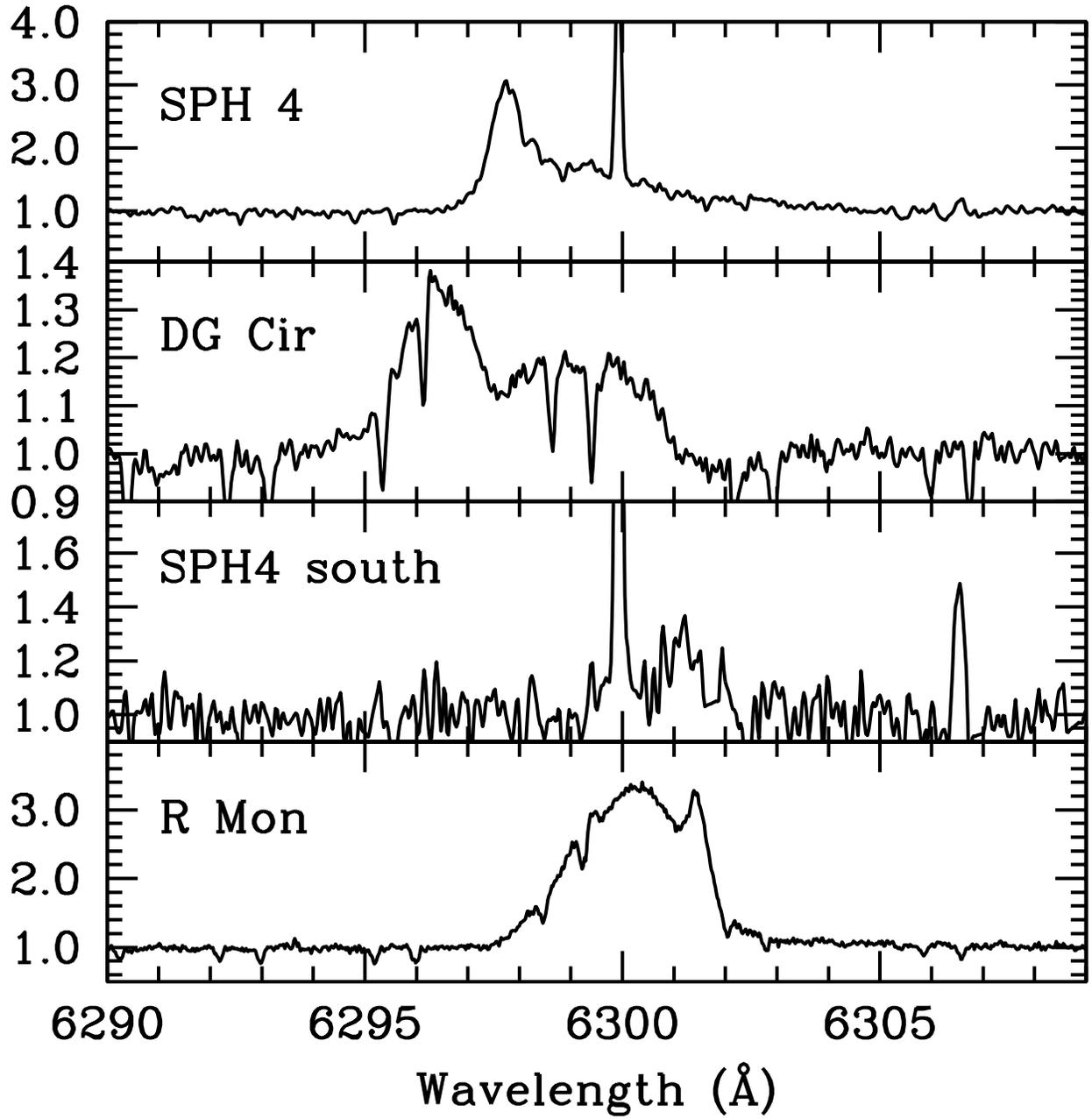}
\caption{Normalized spectra of SPH\,4, DG\,Cir, SPH\,4$-$South, and R\,Mon around the [O\,{\sc i}] line at 6300\,{\AA}.}
\end{figure}

\begin{figure}
\includegraphics[width=\columnwidth]{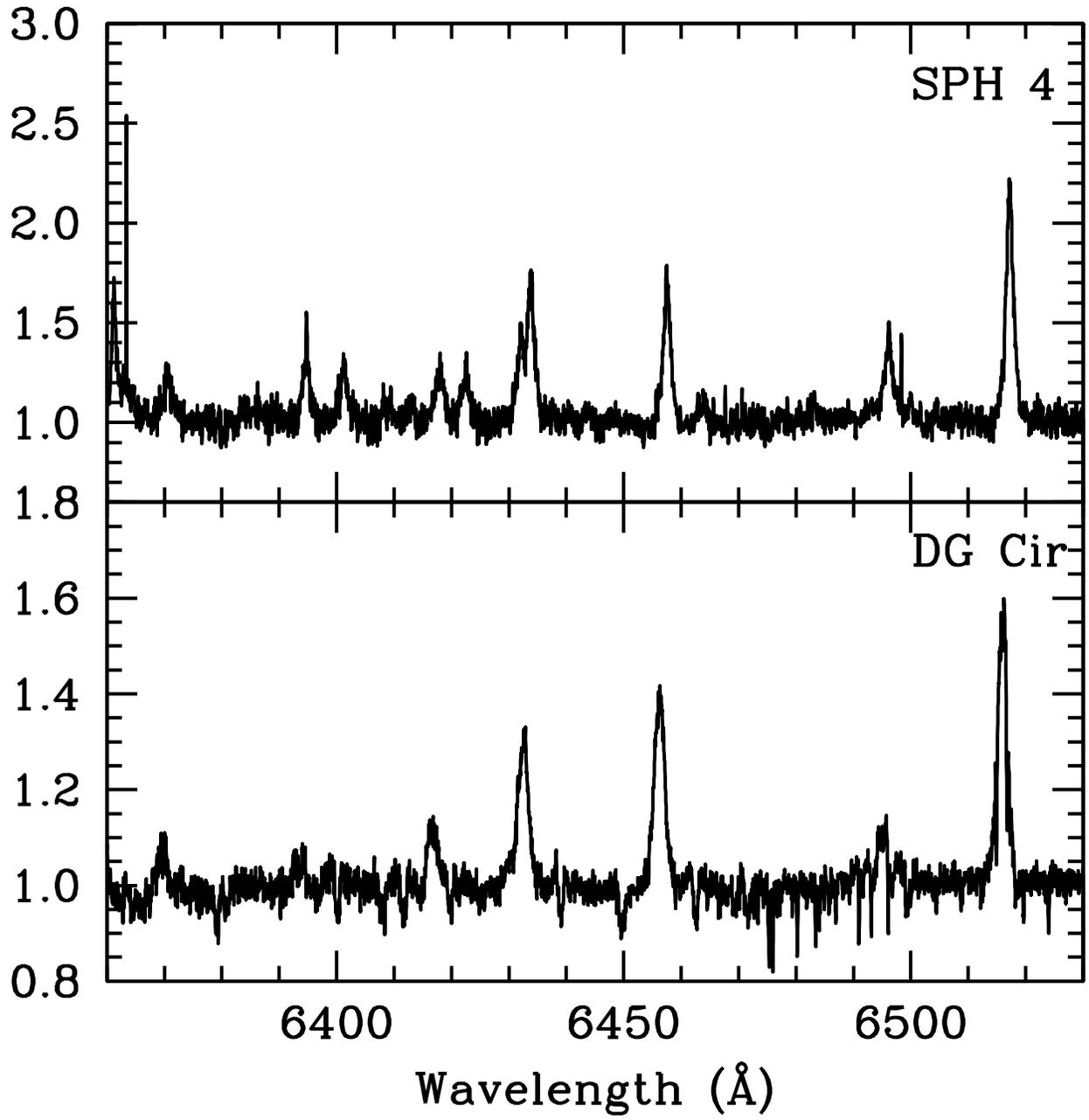}
\caption{Normalized spectra of SPH\,4 and DG\,Cir between 6360 and 6530\,{\AA}.}
\end{figure}

\begin{figure}
\includegraphics[width=\columnwidth]{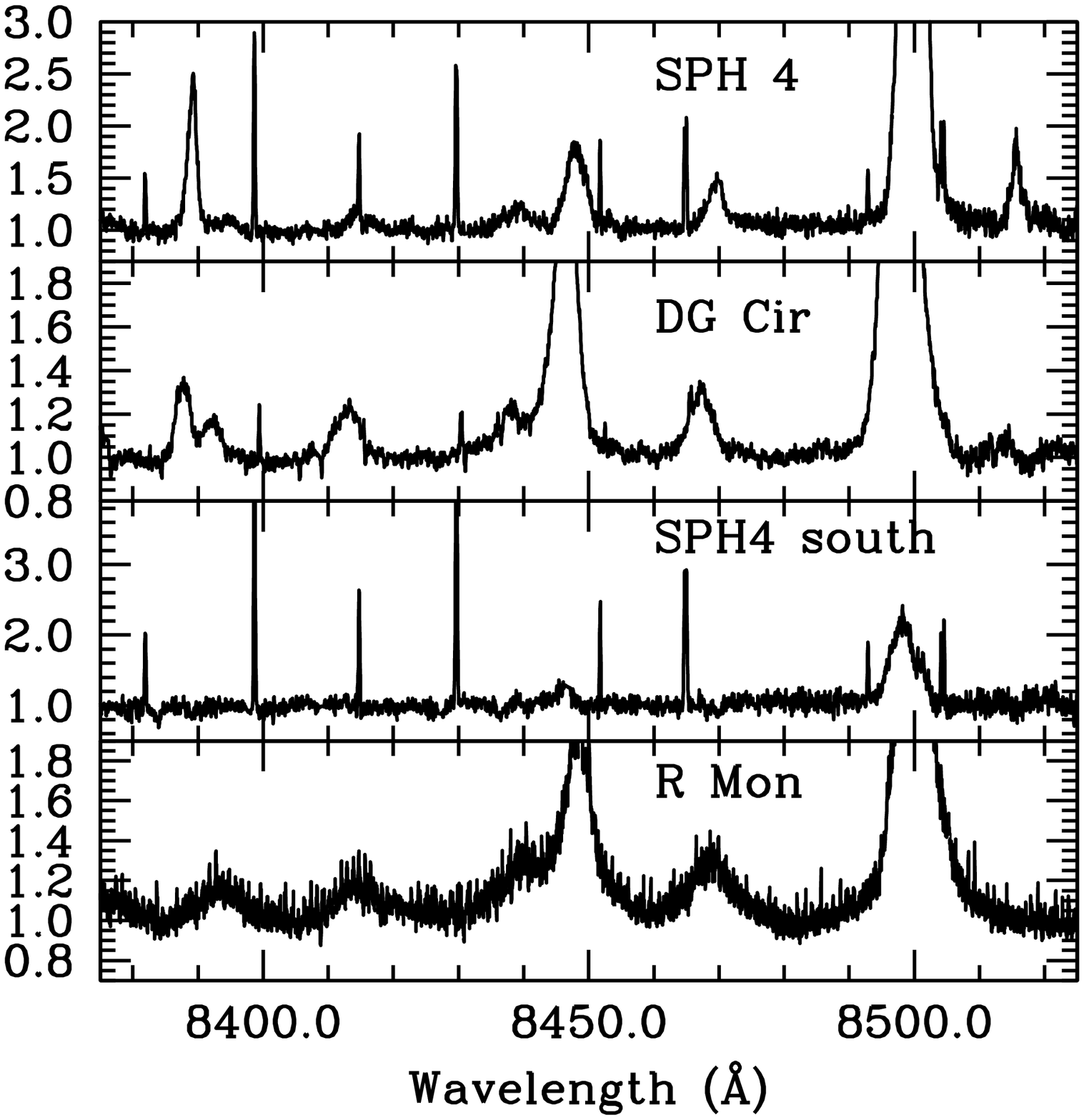}
\caption{Normalized spectra of SPH\,4, DG\,Cir, R\,Mon, and SPH\,4$-$South between 8370 and 8525\,{\AA}. Narrow OH
  emission lines are seen in the spectra of SPH\,4, DG\,Cir, and SPH\,4$-$South.}
\end{figure}

\begin{figure}
\includegraphics[width=\columnwidth]{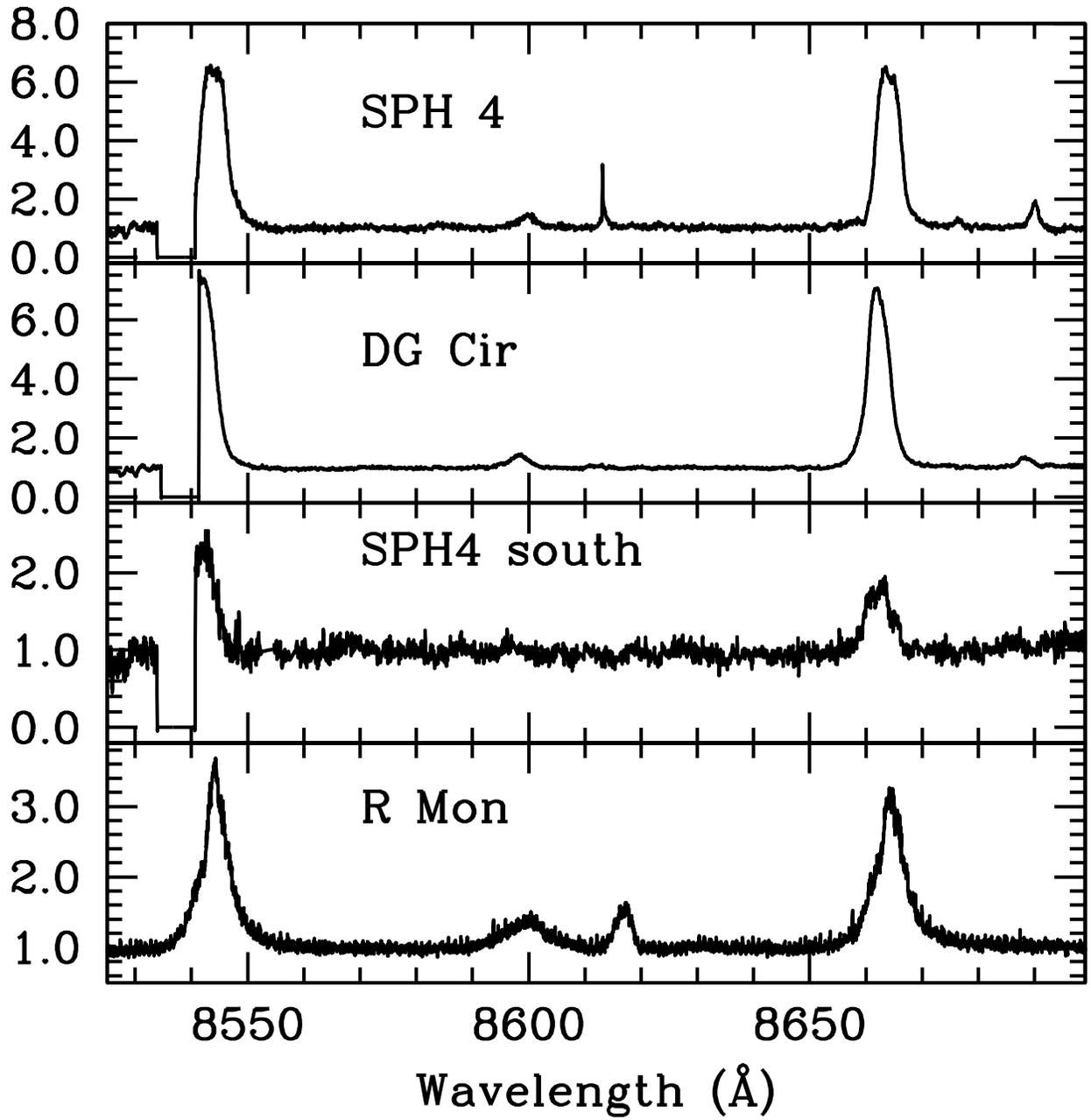}
\caption{Normalized spectra of SPH\,4, DG\,Cir, R\,Mon, and SPH\,4$-$South between 8525 and 8700\,{\AA}.}
\end{figure}

\begin{figure}
\includegraphics[width=\columnwidth]{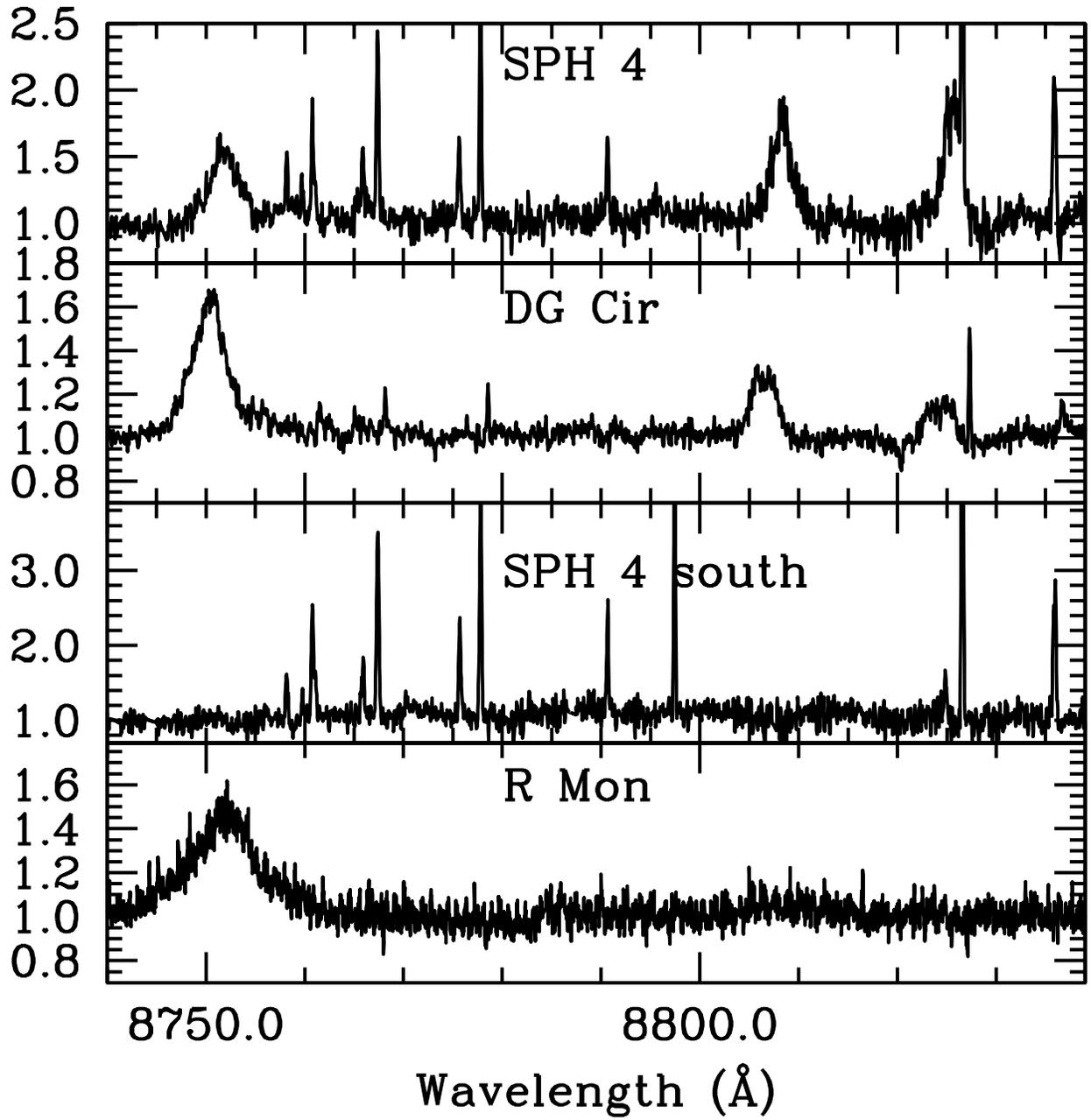}
\caption{Normalized spectra of SPH\,4, DG\,Cir, R\,Mon, and SPH\,4$-$South between 8740 and 8840\,{\AA}.
  Narrow OH emission lines are seen
  in the spectra of SPH\,4, DG\,Cir, and SPH\,4$-$South. }
\end{figure}

\clearpage

\section{Tables}

This section provides measurements of observed wavelength, FWHM,
equivalent width (W$_{\lambda}$), and identification of the emission
lines for the stars studied in this work. The number of the figure
where the corresponding spectrum is included in the table.

\begin{table*}
  \caption{Object, observed wavelength, FWHM, equivalent width (W$_{\lambda}$), and identification of the emission lines for the
    stars studied in this work. Comments are provided for emission lines that present an asymmetric profile, a
    double peak, a flat topped profile, or are contaminated by telluric lines. A few absorption lines were identified and are
    noticed.} 
\begin{tabular}{|l|c|c|c|c|c|}\hline

Star         &  Observed wavelength &  FWHM       & W$_{\lambda}$   &      Line         & Figure  \\
             &   (\AA)              & (\AA)       &     (\AA)     &  Identification   &         \\
\hline
SPH\,4       &   4924.84         &   1.9       &       3.3         &Fe\,{\sc ii}(42) 4923.9 &  10  \\
DG Cir       &   4823.87         &   1.9       &       2.1         &                   &         \\
SPH\,4south  &    ---            &   ---       &       ---         &                   &         \\
R Mon        &    ---            &   ---       &       ---         &                   &         \\
             &  ---------        & ---------   & ---------         & ---------         &         \\
SPH\,4       &    4994.65        &   1.3       &       0.8         &Fe\,{\sc ii}(36) 4993.4 & 10 \\
DG Cir       &                   &             &                   &                   &         \\
SPH\,4south  &    ---            &   ---       &       ---         &                   &         \\
R Mon        &    ---            &   ---       &       ---         &                   &         \\
             &  ---------        & ---------   & ---------         & ---------         &         \\
SPH\,4       &    5012.98        &    1.1      &       0.8         &                   &     11    \\
DG Cir       &    ---            &   ---       &       ---         &                   &         \\
SPH\,4south  &    ---            &   ---       &       ---         &                   &         \\
R Mon        &    ---            &   ---       &       ---         &                   &         \\
             &  ---------        & ---------   & ---------         & ---------         &         \\
SPH\,4       &    5019.43        &    2.2      &       2.4         &Fe\,{\sc ii}(42) 5018.5  & 11   \\
DG Cir       &    5018.53        &    2.1      &       2.6         &                   &         \\
SPH\,4south  &    ---            &   ---       &       ---         &                   &         \\
R Mon        &    ---            &   ---       &       ---         &                   &         \\
             &  ---------        & ---------   & ---------         & ---------         &         \\
SPH\,4       &    5042.33        &    1.7      &       1.3         &                   &    11     \\
DG Cir       &    ---            &   ---       &       ---         &                   &         \\
SPH\,4south  &    ---            &   ---       &       ---         &                   &         \\
R Mon        &    ---            &   ---       &       ---         &                   &         \\
             &  ---------        & ---------   & ---------         & ---------         &         \\
SPH\,4       &    5050.60        &   0.8       &       0.4         &                   &    11   \\
DG Cir       &    ---            &   ---       &       ---         &                   &         \\
SPH\,4south  &    ---            &   ---       &       ---         &                   &         \\
R Mon        &    ---            &   ---       &       ---         &                   &         \\
             &  ---------        & ---------   & ---------         & ---------         &         \\
SPH\,4       &    5052.51        &    1.2      &       0.7         &                   &    11      \\
DG Cir       &    ---            &   ---       &       ---         &                   &         \\
SPH\,4south  &    ---            &   ---       &       ---         &                   &         \\
R Mon        &    ---            &   ---       &       ---         &                   &         \\
             &  ---------        & ---------   & ---------         & ---------         &         \\
SPH\,4       &    5080.42        &    1.7      &       1.2         &                   &     11    \\
DG Cir       &    ---            &   ---       &       ---         &                   &         \\
SPH\,4south  &    ---            &   ---       &       ---         &                   &         \\
R Mon        &    ---            &   ---       &       ---         &                   &         \\
\hline
\end{tabular}
\end{table*}

\begin{table*}
\noindent{Table\,5, continued} 

\begin{tabular}{|l|c|c|c|c|c|}
\hline
Star         &  Observed wavelength &  FWHM    & W$_{\lambda}$    &      Line         & Figure  \\
             &  (\AA)               &   (\AA)     &  (\AA)            &  Identification   &   \\
\hline
SPH\,4       &    5084.23        &    1.3      &       0.5         &                   &    11     \\
DG Cir       &    ---            &   ---       &       ---         &                   &         \\
SPH\,4south  &    ---            &   ---       &       ---         &                   &         \\
R Mon        &    ---            &   ---       &       ---         &                   &         \\
             &  ---------        & ---------   & ---------         & ---------         &         \\
SPH\,4       &    5108.45        &   1.3       &       0.8         &                   &    12   \\
DG Cir       &    ---            &   ---       &       ---         &                   &         \\
SPH\,4south  &    ---            &   ---       &       ---         &                   &         \\
R Mon        &    ---            &   ---       &       ---         &                   &         \\
             &  ---------        & ---------   & ---------         & ---------         &         \\
SPH\,4       &    5111.18        &   1.3       &       0.9         &                   &    12     \\
DG Cir       &    5110.51        &   1.5       &       0.3         &                   &         \\
SPH\,4south  &    ---            &   ---       &       ---         &                   &         \\
R Mon        &    ---            &   ---       &       ---         &                   &         \\
             &  ---------        & ---------   & ---------         & ---------         &         \\
SPH\,4       &    5128.27        &   1.0       &       0.3         &                   &    12     \\
DG Cir       &    5128.72        &   2.3       &       0.1         &                   &         \\
SPH\,4south  &    ---            &   ---       &       ---         &                   &         \\
R Mon        &    ---            &   ---       &       ---         &                   &         \\
             &  ---------        & ---------   & ---------         & ---------         &         \\
SPH\,4       &    5129.97        &   1.1       &       0.5         &                   &    12    \\
DG Cir       &    ---            &   ---       &       ---         &                   &         \\
SPH\,4south  &    ---            &   ---       &       ---         &                   &         \\
R Mon        &    ---            &   ---       &       ---         &                   &         \\
             &  ---------        & ---------   & ---------         & ---------         &         \\
SPH\,4       &    5152.05        &   1.4       &       0.6         &                   &    12     \\
DG Cir       &    ---            &   ---       &       ---         &                   &         \\
SPH\,4south  &    ---            &   ---       &       ---         &                   &         \\
R Mon        &    ---            &   ---       &       ---         &                   &         \\
             &  ---------        & ---------   & ---------         & ---------         &         \\
SPH\,4       &    5156.80        &   1.1       &       0.8         &                   &    12      \\
DG Cir       &    ---            &   ---       &       ---         &                   &         \\
SPH\,4south  &    ---            &   ---       &       ---         &                   &         \\
R Mon        &    ---            &   ---       &       ---         &                   &         \\
             &  ---------        & ---------   & ---------         & ---------         &         \\
SPH\,4       &    5169.80        &   2.2       &       2.3         & Fe\,{\sc ii}(42) 5169.0  & 12 \\
DG Cir       &    5168.92        &   3.0       &       3.1         &                   &         \\
SPH\,4south  &    ---            &   ---       &       ---         &                   &         \\
R Mon        &    ---            &   ---       &       ---         &                   &         \\
             &  ---------        & ---------   & ---------         & ---------         &         \\
SPH\,4       &    5173.61        &   2.1       &       2.1         & Mg\,{\sc i} 5172.7 &   12     \\
DG Cir       &    5172.51        &   1.6       &       1.5         &                   &         \\
SPH\,4south  &    ---            &   ---       &       ---         &                   &         \\
R Mon        &    ---            &   ---       &       ---         &                   &         \\
\hline
\end{tabular}
\end{table*}

\begin{table*}
\noindent{Table\,5, continued} 

\begin{tabular}{|l|c|c|c|c|c|}
\hline
Star         &  Observed wavelength &  FWHM       & W$_{\lambda}$    &      Line         & Figure  \\
             &  (\AA)               &   (\AA)     &  (\AA)         &  Identification   &         \\\hline
SPH\,4       &    5173.61        &   2.1       &       2.1         & Mg\,{\sc i} 5172.7 &  12    \\
DG Cir       &    5172.51        &   1.6       &       1.5         &                   &         \\
SPH\,4south  &    ---            &   ---       &       ---         &                   &         \\
R Mon        &    ---            &   ---       &       ---         &                   &         \\
             &  ---------        & ---------   & ---------         & ---------         &         \\
SPH\,4       &    5184.50        &   ---       &      ---          & Mg\,{\sc i} 5183.6 &  12    \\
             &   asymmetric      &             &                   &                   &         \\
             &    profile        &             &                   &                   &         \\
DG Cir       &    5183.60        &    2.5      &       1.3         &                   &         \\
SPH\,4south  &    ---            &   ---       &       ---         &                   &         \\
R Mon        &    ---            &   ---       &       ---         &                   &         \\
             &  ---------        & ---------   & ---------         & ---------         &         \\
SPH\,4       &    5198.50        &    1.6      &       1.4         & Fe\,{\sc ii} (49) 5197.6  & 12 \\
DG Cir       &    5197.52        &    1.5      &       0.9         &                   &         \\
SPH\,4south  &    ---            &   ---       &       ---         &                   &         \\
R Mon        &    ---            &   ---       &       ---         &                   &         \\
             &  ---------        & ---------   & ---------         & ---------         &         \\
SPH\,4       &    5217.17        &   ---       &       ---         &                   &    13     \\
             & broad double peak &             &                   &                   &         \\
DG Cir       &    ---            &   ---       &       ---         &                   &         \\
SPH\,4south  &    ---            &   ---       &       ---         &                   &         \\
R Mon        &    ---            &   ---       &       ---         &                   &         \\
             &  ---------        & ---------   & ---------         & ---------         &         \\
SPH\,4       &    5227.95        &   1.7       &       1.2         &                   &   13      \\
DG Cir       &    5226.72        &   2.4       &       0.7         &                   &         \\
SPH\,4south  &    ---            &   ---       &       ---         &                   &         \\
R Mon        &    ---            &   ---       &       ---         &                   &         \\
             &  ---------        & ---------   & ---------         & ---------         &         \\
SPH\,4       &    5233.92        &   0.9       &       0.3         &                   &   13     \\
DG Cir       &    ---            &   ---       &       ---         &                   &         \\
SPH\,4south  &    ---            &   ---       &       ---         &                   &         \\
R Mon        &    ---            &   ---       &       ---         &                   &         \\
             &  ---------        & ---------   & ---------         & ---------         &         \\
SPH\,4       &    5235.51        &   2.0       &       1.6         & Fe\,{\sc ii}(49) 5234.6   & 13 \\
DG Cir       &    5234.48        &   2.2       &       0.9         &                   &         \\
SPH\,4south  &    ---            &   ---       &       ---         &                   &         \\
R Mon        &    ---            &   ---       &       ---         &                   &         \\
             &  ---------        & ---------   & ---------         & ---------         &         \\
SPH\,4       &    5270.78        &   1.9       &       2.1         &                   &   13      \\
DG Cir       &    5259.58        &   2.0       &       1.2         &                   &         \\
SPH\,4south  &    ---            &   ---       &       ---         &                   &         \\
R Mon        &    ---            &   ---       &       ---         &                   &         \\
\hline
\end{tabular}
\end{table*}

\begin{table*}
\noindent{Table\,5, continued} 

\begin{tabular}{|l|c|c|c|c|c|}
\hline
Star         &  Observed wavelength &  FWHM       & W$_{\lambda}$    &      Line         & Figure  \\
             &  (\AA)               &   (\AA)     &  (\AA)         &  Identification   &         \\
\hline

SPH\,4       &    5276.95        &   1.5       &       2.2         &Fe\,{\sc ii}(49) 5276.0   & 13 \\
DG Cir       &    5275.95        &   2.0       &       1.5         &                   &         \\
SPH\,4south  &    ---            &   ---       &       ---         &                   &         \\
R Mon        &    ---            &   ---       &       ---         &                   &         \\
             &  ---------        & ---------   & ---------         & ---------         &         \\
SPH\,4       &    5283.89        &   ---       &       ---         & Fe\,{\sc ii}(41) 5284.1 & 13  \\
             & broad double peak &             &                   &                   &         \\
DG Cir       &    5285.00        &   1.1       &       1.2         &                   &         \\
SPH\,4south  &    ---            &   ---       &       ---         &                   &         \\
R Mon        &    ---            &   ---       &       ---         &                   &         \\
             & ---------         & ---------   & ---------         & ---------         &         \\
SPH\,4       &   5317.58         &   ---       &       ---         & Fe\,{\sc ii}(49) 5316.2 & 14  \\
DG Cir       &   5316.58         &   ---       &       ---         &                   &         \\
SPH\,4south  &    ---            &   ---       &       ---         &                   &         \\
R Mon        &    ---            &   ---       &       ---         &                   &         \\
             &  ---------        & ---------   & ---------         & ---------         &         \\
SPH\,4       &    5363.80        &   1.6       &       1.5         &Fe\,{\sc ii}(48) 5362.9  & 14  \\
DG Cir       &    5362.75        &   1.7       &       0.8         &                   &         \\
SPH\,4south  &    ---            &   ---       &       ---         &                   &         \\
R Mon        &    ---            &   ---       &       ---         &                   &         \\
             &  ---------        & ---------   & ---------         & ---------         &         \\
SPH\,4       &    5372.34        &   1.0       &       0.8         &                   &    14   \\
DG Cir       &    5371.22        &   2.7       &       0.4         &                   &         \\
             &flat topped profile&             &                   &                   &         \\
SPH\,4south  &    ---            &   ---       &       ---         &                   &         \\
R Mon        &    ---            &   ---       &       ---         &                   &         \\
             &  ---------        & ---------   & ---------         & ---------         &         \\
SPH\,4       &    5398.00        &   1.2       &       1.2         &                   &    14      \\
DG Cir       &    5396.94        &   1.6       &       0.4         &                   &         \\
SPH\,4south  &    ---            &   ---       &       ---         &                   &         \\
R Mon        &    ---            &   ---       &       ---         &                   &         \\
             &  ---------        & ---------   & ---------         & ---------         &         \\
SPH\,4       &    5406.70        &   1.6       &       0.9         &                   &    15      \\
DG Cir       &    ---            &   ---       &       ---         &                   &         \\
SPH\,4south  &    ---            &   ---       &       ---         &                   &         \\
R Mon        &    ---            &   ---       &       ---         &                   &         \\
             &  ---------        & ---------   & ---------         & ---------         &         \\
SPH\,4       &    5415.35        &   2.1       &       0.6         &                   &    15    \\
DG Cir       &    ---            &   ---       &       ---         &                   &         \\
SPH\,4south  &    ---            &   ---       &       ---         &                   &         \\
R Mon        &    ---            &   ---       &       ---         &                   &         \\
\hline
\end{tabular}
\end{table*}

\begin{table*}
\noindent{Table\,5, continued} 

\begin{tabular}{|l|c|c|c|c|c|}
\hline
Star         &  Observed wavelength &  FWHM    & W$_{\lambda}$      &      Line         & Figure  \\
             &  (\AA)               &   (\AA)     &  (\AA)        &  Identification   &         \\
\hline
SPH\,4       &    5425.92        &   ---       &       ---         & Fe\,{\sc ii}(49) 5425.3 &   \\
            &asymmetric profile  &             &                   &                   &    15     \\
DG Cir       &     5425.24       &   ---       &       ---         &                   &         \\
            &flat topped profile &             &                   &                   &         \\
SPH\,4south  &    ---            &   ---       &       ---         &                   &         \\
R Mon        &    ---            &   ---       &       ---         &                   &         \\
             &  ---------        & ---------   & ---------         & ---------         &         \\
SPH\,4       &    5430.60        &   1.4       &       1.2         &                   &    15      \\
DG Cir       &    5429.53        &   2.0       &       0.4         &                   &         \\
SPH\,4south  &    ---            &   ---       &       ---         &                   &         \\
R Mon        &    ---            &   ---       &       ---         &                   &         \\
             &  ---------        & ---------   & ---------         & ---------         &         \\
SPH\,4       &    5435.09        &   ---       &       ---         &                   &    15      \\
            &asymmetric profile  &             &                   &                   &         \\
DG Cir       &    5434.08        &   ---       &       ---         &                   &         \\
             &   double peak     &             &                   &                   &         \\
SPH\,4south  &    ---            &   ---       &       ---         &                   &         \\
R Mon        &    ---            &   ---       &       ---         &                   &         \\
             &  ---------        & ---------   & ---------         & ---------         &         \\
SPH\,4       &    5447.83        &   1.2       &       0.8         &                   &     15     \\
DG Cir       &    5447.85        &   2.0       &       0.4         &                   &         \\
             &   double peak     &             &                   &                   &         \\
SPH\,4south  &    ---            &   ---       &       ---         &                   &         \\
R Mon        &    ---            &   ---       &       ---         &                   &         \\
             &  ---------        & ---------   & ---------         & ---------         &         \\
SPH\,4       &    5456.50        &   1.2       &       0.8         &                   &     15      \\
DG Cir       &    ---            &   ---       &       ---         &                   &         \\
SPH\,4south  &    ---            &   ---       &       ---         &                   &         \\
R Mon        &    ---            &   ---       &       ---         &                   &         \\
             &  ---------        & ---------   & ---------         & ---------         &         \\
SPH\,4       &    6138.27        &   2.1       &       1.6         &                   &     17     \\
DG Cir       &    6137.18        &   2.0       &       0.4         &                   &         \\
             &  double peak      &             &                   &                   &         \\
SPH\,4south  &    ---            &   ---       &       ---         &                   &         \\
R Mon        &    ---            &   ---       &       ---         &                   &         \\
             &  ---------        & ---------   & ---------         & ---------         &         \\
SPH\,4       &    6142.45        &   1.4       &       0.3         &                   &     17      \\
DG Cir       &    ---            &   ---       &       ---         &                   &         \\
SPH\,4south  &    ---            &   ---       &       ---         &                   &         \\
R Mon        &    ---            &   ---       &       ---         &                   &         \\
\hline
\end{tabular}
\end{table*}

\begin{table*}
\noindent{Table\,5, continued} 

\begin{tabular}{|l|c|c|c|c|c|}
\hline
Star         &  Observed wavelength &  FWHM    & W$_{\lambda}$      &      Line         & Figure  \\
             &  (\AA)               &   (\AA)     &  (\AA)        &  Identification   &         \\
\hline
SPH\,4       &    6149.68        &   ---       &       ---         & Fe\,{\sc ii}(74) 6149.2 & 17    \\
             &  double peak      &             &                   &                   &         \\
DG Cir       &    6148.35        &   3.1       &       0.5         &                   &         \\
SPH\,4south  &    ---            &   ---       &       ---         &                   &         \\
R Mon        &    6150.0         &   5.0       &       1.6         &                   &        \\
             &  ---------        & ---------   & ---------         & ---------         &         \\
SPH\,4       &    6192.53        &   1.6       &       0.9         &                   &    17      \\
DG Cir       &    ---            &   ---       &       ---         &                   &         \\
SPH\,4south  &    ---            &   ---       &       ---         &                   &         \\
R Mon        &    ---            &   ---       &       ---         &                   &         \\
             &  ---------        & ---------   & ---------         & ---------         &         \\
SPH\,4       &    6231.81        &   1.8       &       0.6         &                   &    17     \\
DG Cir       &    ---            &   ---       &       ---         &                   &         \\
SPH\,4south  &    ---            &   ---       &       ---         &                   &         \\
R Mon        &    ---            &   ---       &       ---         &                   &         \\
             &  ---------        & ---------   & ---------         & ---------         &         \\
SPH\,4       &    6239.64        &   2.0       &       0.7         & Fe\,{\sc ii}(34) 6238.3 & 17  \\
DG Cir       &    6238.56        &   ---       &       ---         &                   &         \\
             &asymmetric profile &             &                   &                   &         \\
SPH\,4south  &    ---            &   ---       &       ---         &                   &         \\
R Mon        &    6240.11        &   4.8       &       0.5         &                   &         \\
             &  ---------        & ---------   & ---------         & ---------         &         \\
SPH\,4       &    6248.52        &   1.7       &       1.1         & Fe\,{\sc ii}(34) 6247.5 & 17  \\
DG Cir       &    6247.45        &   2.1       &       0.6         &                   &         \\
SPH\,4south  &    ---            &   ---       &       ---         &                   &         \\
R Mon        &    6249.00        &   4.5       &       1.0         &                   &         \\
             &  ---------        & ---------   & ---------         & ---------         &         \\
SPH\,4       &    6253.69        &   1.3       &       0.5         &                   &    17     \\
DG Cir       &    ---            &   ---       &       ---         &                   &         \\
SPH\,4south  &    ---            &   ---       &       ---         &                   &         \\
R Mon        &    ---            &   ---       &       ---         &                   &         \\
             &  ---------        & ---------   & ---------         & ---------         &         \\
SPH\,4       &    6394.71        &   1.6       &       0.6         &                   &    19     \\
DG Cir       &    ---            &   ---       &       ---         &                   &         \\
SPH\,4south  &    ---            &   ---       &       ---         &                   &         \\
R Mon        &    ---            &   ---       &       ---         &                   &         \\
             &  ---------        & ---------   & ---------         & ---------         &         \\
SPH\,4       &    6401.19        &   1.7       &       0.4         &                   &    19   \\
DG Cir       &    ---            &   ---       &       ---         &                   &         \\
SPH\,4south  &    ---            &   ---       &       ---         &                   &         \\
R Mon        &    ---            &   ---       &       ---         &                   &         \\
\hline
\end{tabular}
\end{table*}

\begin{table*}
\noindent{Table\,5, continued} 

\begin{tabular}{|l|c|c|c|c|c|}
\hline
Star         &  Observed wavelength &  FWHM    & W$_{\lambda}$      &      Line         & Figure  \\
             &  (\AA)               &   (\AA)     &  (\AA)        &  Identification   &         \\
\hline
SPH\,4       &    6417.98        &   2.3       &       0.5         & Fe\,{\sc ii}(74) 6416.9  & 19 \\
DG Cir       &    6416.71        &   ---       &       ---         &                   &         \\
             &asymmetric profile &             &                   &                   &         \\
SPH\,4south  &    ---            &   ---       &       ---         &                   &         \\
R Mon        &    ---            &   ---       &       ---         &                   &         \\
             &  ---------        & ---------   & ---------         & ---------         &         \\
SPH\,4       &    6422.44        &   1.9       &       0.4         &                   &    19     \\
DG Cir       &    ---            &   ---       &       ---         &                   &         \\
SPH\,4south  &    ---            &   ---       &       ---         &                   &         \\
R Mon        &    ---            &   ---       &       ---         &                   &         \\
             &  ---------        & ---------   & ---------         & ---------         &         \\
SPH\,4       &    6432.40        &   1.9       &       0.4         & Fe\,{\sc ii}(40) 6432.7 & 19  \\
             &  double peak      &             &                   &                   &         \\
DG Cir       &    6431.48        &   ---       &       ---         &                   &         \\
             &asymmetric profile &             &                   &                   &         \\
SPH\,4south  &    ---            &   ---       &       ---         &                   &         \\
R Mon        &    ---            &   ---       &       ---         &                   &         \\
             &  ---------        & ---------   & ---------         & ---------         &         \\
SPH\,4       &    6457.53        &   1.8       &       1.3         & Fe\,{\sc ii}(74) 6456.4 & 19  \\
DG Cir       &    6456.23        &   2.1       &       0.9         &                   &         \\
SPH\,4south  &    ---            &   ---       &       ---         &                   &         \\
R Mon        &    ---            &   ---       &       ---         &                   &         \\
             &  ---------        & ---------   & ---------         & ---------         &         \\
SPH\,4       &    6496.15        &   1.7       &       0.7         &                   &    19     \\
DG Cir       &    ---            &   ---       &       ---         &                   &         \\
SPH\,4south  &    ---            &   ---       &       ---         &                   &         \\
R Mon        &    ---            &   ---       &       ---         &                   &         \\
             &  ---------        & ---------   & ---------         & ---------         &         \\
SPH\,4       &    6517.23        &   1.6       &       1.8         & Fe\,{\sc ii}(40) 6516.1 & 19  \\
DG Cir       &    6515.93        &   1.8       &       1.1         &                   &         \\
SPH\,4south  &    ---            &   ---       &       ---         &                   &         \\
R Mon        &    ---            &   ---       &       ---         &                   &         \\
             &  ---------        & ---------   & ---------         & ---------         &         \\

SPH\,4       &    8348.97        &   5.8       &       1.3         &                   &    20     \\
DG Cir       &    ---            &   ---       &       ---         &                   &         \\
SPH\,4south  &    ---            &   ---       &       ---         &                   &         \\
R Mon        &    ---            &   ---       &       ---         &                   &         \\
             &  ---------        & ---------   & ---------         & ---------         &         \\
SPH\,4       &    8389.13        &   1.6       &       2.2         &                   &   20      \\
DG Cir       &    8387.69        &   2.6       &       1.0         &                   &         \\
SPH\,4south  &    ---            &   ---       &       ---         &                   &         \\
R Mon        &    ---            &   ---       &       ---         &                   &         \\
\hline
\end{tabular}
\end{table*}

\begin{table*}
\noindent{Table\,5, continued} 

\begin{tabular}{|l|c|c|c|c|c|}
\hline
Star         &  Observed wavelength &  FWHM    & W$_{\lambda}$      &      Line         & Figure  \\
             &  (\AA)               &   (\AA)     &  (\AA)        &  Identification   &         \\
\hline
SPH\,4       &    8394.65        &   3.5       &       0.6         & H\,{\sc i}(P20) 8392.40 & 20   \\
DG Cir       &    8392.05        &   ---       &       ---         &                   &         \\
             &asymmetric profile &             &                   &                   &         \\
SPH\,4south  &    ---            &   ---       &       ---         &                   &         \\
R Mon        &    8383.60        &   9.5       &       1.8         &                   &         \\
             &  ---------        & ---------   & ---------         & ---------         &         \\
SPH\,4       &    8414.73        &   ---       &       ---         & H\,{\sc i}(Pa19) 8413.30 & 20  \\
             &contaminated by OH emission &    &                   &                          &  \\
DG Cir       &    8412.93        &   4.2       &       1.0         &                   &         \\
SPH\,4south  &    ---            &   ---       &       ---         &                   &         \\
R Mon        &    8415.26        &   10.9      &       2.3         &                   &         \\
             &  ---------        & ---------   & ---------         & ---------         &         \\
SPH\,4       &    8439.35        &   5.4       &       1.0         &                   &   20      \\
DG Cir       &    8438.18        &   4.4       &       1.1         &                   &         \\
SPH\,4south  &    ---            &   ---       &       ---         &                   &         \\
R Mon        &    8440.38        &   9.1       &       2.4         &                   &         \\
             &  ---------        & ---------   & ---------         & ---------         &         \\
SPH\,4       &    8448.01        &   3.5       &       3.0         & O\,{\sc i} 8446.74&   20      \\
DG Cir       &    8446.43        &   3.8       &       4.3         &                   &         \\
SPH\,4south  &    8446.19        &   3.0       &       0.6         &                   &         \\
             &    double peak    &             &                   &                   &         \\
R Mon        &    8448.52        &   5.3       &       4.5         &                   &         \\
&  ---------        & ---------   & ---------         & ---------         &         \\
SPH\,4       &    8469.58        &   2.9       &       1.0         & H\,{\sc i}(Pa17) 8467.80  & 20  \\
DG Cir       &    8467.22        &   4.0       &       1.3         &                   &         \\
SPH\,4south  &    ---            &   ---       &       ---         &                   &         \\
R Mon        &    8468.84        &   7.9       &       2.0         &                   &         \\
             &  ---------        & ---------   & ---------         & ---------         &         \\
SPH\,4       &    8499.67        &   3.9       &      27.4         & Ca\,{\sc ii} 8498.06  & 20      \\
DG Cir       &    8498.02        &   3.6       &      30.0         &                   &         \\
SPH\,4south  &    8498.17        &   5.5       &       5.4         &                   &         \\
R Mon        &    8499.98        &   5.0       &      14.5         &                   &         \\
             &  ---------        & ---------   & ---------         & ---------         &         \\
SPH\,4       &    8515.64        &   1.8       &       1.1        & Ca\,{\sc ii} 8498.06   & 20    \\
DG Cir       &    ---            &   ---       &      ---         &                   &         \\
SPH\,4south  &    ---            &   ---       &      ---         &                   &         \\
R Mon        &    ---            &   ---       &      ---         &                   &         \\
             &  ---------        & ---------   & ---------         & ---------         &         \\
SPH\,4       &    ---            &   ---       &       ---         &  Ca\,{\sc ii} 8542.14&  21    \\
DG Cir       &    ---            &   ---       &       ---         &                   &         \\
SPH\,4south  &    ---            &   ---       &       ---         &                   &         \\
R Mon        &   8544.31         &   6.0       &       12.8        &                   &         \\
\hline
\end{tabular}
\end{table*}

\begin{table*}
\noindent{Table\,5, continued} 

\begin{tabular}{|l|c|c|c|c|c|}
 \hline
Star         &  Observed wavelength &  FWHM    & W$_{\lambda}$    &      Line         & Figure  \\
             &  (\AA)               &   (\AA)     &  (\AA)            &  Identification   &         \\
\hline 
SPH\,4       &   8599.73         &   3.7       &       1.6         & H\,{\sc i} 8598.39&    21     \\
DG Cir       &   8598.28         &   4.3       &       4.3         &                   &         \\
SPH\,4south  &    ---            &   ---       &       ---         &                   &         \\
R Mon        &   8599.60         &   10.0      &       3.4         &                   &         \\
&  --------         & ---------   & ---------         & ---------         &         \\
SPH\,4       &   ---             &   ---       &       ---         &                   &    21     \\
DG Cir       &   ---             &   ---       &       ---         &                   &         \\
SPH\,4south  &    ---            &   ---       &       ---         &                   &         \\
R Mon        &   8616.92         &   3.3       &       1.9         &                   &         \\
             &  ---------        & ---------   & ---------         & ---------         &         \\
SPH\,4       &   8664.00         &   4.4       &      21.9         & Ca\,{\sc ii} 8662.17 &   21   \\
DG Cir       &   8662.20         &   4.5       &      29.6         &                   &         \\
SPH\,4south  &   8662.32         &   5.3       &       4.4         &                   &         \\
R Mon        &   8664.38         &   6.2       &      11.9         &                   &         \\
             &  ---------        & ---------   & ---------         & ---------         &         \\
SPH\,4       &   8690.01         &   1.8       &       1.5         &                   &    22     \\
DG Cir       &   8688.38         &   2.4       &       0.7         &                   &         \\
SPH\,4south  &   ---             &   ---       &      ---          &                   &         \\
R Mon        &   ---             &   ---       &      ---          &                   &         \\
& ---------         & ---------   & ---------         & ---------         &         \\
SPH\,4       &   8751.90         &   3.5       &      1.6          &H\,{\sc i}(Pa12) 8750.38&  22  \\
DG Cir       &   8750.34         &   4.2       &      2.6          &                   &         \\
SPH\,4south  &   ---             &   ---       &      ---          &                   &         \\
R Mon        &   8751.83         &   9.5       &      4.1          &                   &         \\
             &  --------         & ---------   & ---------         & ---------         &         \\
SPH\,4       &   8808.32         &   2.4       &      1.6          &                   &   22      \\
DG Cir       &   8806.44         &   2.9       &      0.9          &                   &         \\
SPH\,4south  &   ---             &   ---       &      ---          &                   &         \\
R Mon        &   ---             &   ---       &      ---          &                   &         \\
&  --------         & ---------   & ---------         & ---------         &         \\
SPH\,4       &   8825.40         &   ---       &      ---          & Fe\,{\sc ii} 8824.59  &  22       \\
             &contaminated by OH emission &    &                   &                   &         \\
DG Cir       &   8824.40         &   3.4       &      0.7          &                   &         \\
SPH\,4south  &   ---             &   ---       &      ---          &                   &         \\
R Mon        &   ---             &   ---       &      ---          &                   &         \\
\hline
\end{tabular}
\end{table*}


\begin{thebibliography}{99}

\bibitem[\protect\citeauthoryear{Brand et al.}{1986}]{b13}Brand J., Blitz L., Wouterloot J.~G.~A.,
1986, A\&ASS, 65, 537

\bibitem[\protect\citeauthoryear{Chambers et al}{2016}]{b15}Chambers K.~C. et al., 2016, arXiv:1612.05560
  
\bibitem[\protect\citeauthoryear{Cutri et al}{2003}]{b17}Cutri, R.M., Skrutski, M.F.,
van Dyk, S., Beichman, C.A., Carpenter, J.M. et al. 2003, The IRSA 2MASS All Sky Point Source
Catalog, NASA/IPAC Infrared Science Archive
  
\bibitem[\protect\citeauthoryear{Dahm}{2005}]{b1}Dahm S.~E, Simon T., 2005, AJ, 129, 829

\bibitem[\protect\citeauthoryear{Fischer et al.}{2016}]{b18}Fischer W.~E., Padgett D.~L., Stapelfeldt K.~L., Sewi{\l}o M., 2016, ApJ, 876, A96

\bibitem[\protect\citeauthoryear{Flewelling et al}{2016}]{b16}Flewelling H.A., et al., 2016, arXiv:1612.05243
  
\bibitem[\protect\citeauthoryear{Gaia2020}{2020}]{b14}Gaia Collaboration, Brown A.G.A., et al., 2020, arXiv:2012.01533 
  
\bibitem[\protect\citeauthoryear{Hanuschik}{2003}]{b2}Hanuschik R.~W., 2003, A\&A, 407, 1157.

\bibitem[\protect\citeauthoryear{Herbig}{2003}]{b3}Herbig G.~H., Petrov P.~P., Duemmler R., 2003, ApJ, 595, 384

\bibitem[\protect\citeauthoryear{Her\'andez}{2004}]{b4}Hern\'andez J., Calvet N., Brice\~no C., Hartmann L., Berlind P., 2004, AJ, 127, 1682

\bibitem[\protect\citeauthoryear{Kaufer et al}{1999}]{b5}Kaufer A., Stahl O., Tubbesing S., N\"orregaard P., Avila G., Francois P., 1999, The Messenger, 95, 8

\bibitem[\protect\citeauthoryear{Lada}{1962}]{b6}Lada C.~J., 1987, in Proc. IAU Symp. 115, Star Forming Regions, ed. M. Peimbert \& J. Jugaku (Dordrecht: Reidel), 1

\bibitem[\protect\citeauthoryear{Lynds}{1962}]{b6}Lynds B.~T., 1962, ApJS, 1962, 7, 1

\bibitem[\protect\citeauthoryear{May et al.}{2005}]{b7}May J., Gyulbudaghian A.~L., Alvarez H., 2005, Astrophysics., 48, 411

\bibitem[\protect\citeauthoryear{Osterbrock et al}{1996}]{b8}Osterbrock D.~E., Fulbright J.~P., Martel A.~R., Keane M.~J., Trager S.~C.,
  Basri G., 1996, PASP, 108, 277

\bibitem[\protect\citeauthoryear{Percy et al}{2010}]{b9}Percy J.~R., Seneviratne R., Herbst W., 2010, PASP, 122 753

\bibitem[\protect\citeauthoryear{Pereira et al}{2001}]{b10}Pereira C.~B., Schiavon R.~P., de Ara\'ujo F.~X., Landaberry S.~J.~C., 2001, AJ, 121, 1071

\bibitem[\protect\citeauthoryear{Schwartz et al}{1990}]{b11}Schwartz R.~D., Persson S.~E., Hamann F.~W., 1990, AJ, 100, 793

\bibitem[\protect\citeauthoryear{Vioque et al}{2018}]{b12}Vioque M., Oudmaijer R.~D. Baines D., Mendigut\'{\i}a I., P\'erez-Mart\'{\i}nez R., 2018, A\&A, 620, A128

\end{thebibliography}
\end{document}